\documentclass[12pt]{iopart}

\usepackage{graphicx}
\usepackage{xcolor}

\begin{document}




\title{Dynamics of phase space vortices in Vlasov plasmas with ion scale inhomogeneity : II Chirped frequency drive study}


\author{Sanjeev Kumar Pandey$^{[1]}$, Amudon Chingangbam$^{[2,3]}$  and Rajaraman Ganesh$^{[2,3]}$}

\address{$^{1}$Department of Physics, Indian Institute of Technology (IIT) Madras, Chennai 600036, India}
\address{$^{2}$Institute for Plasma Research (IPR), Bhat, Gandhinagar 382428, India}
\address{$^{3}$Homi Bhabha National Institute (HBNI), Mumbai, Maharashtra 400094, India}
\ead{sanju23510@gmail.com and ganesh@ipr.res.in}
\vspace{10pt}
\begin{indented}
   \item July 2026
\end{indented}

\begin{abstract}
In Part I of the companion paper [--REF PART i--], we have extensively discussed about the creation of quasi-stationary ion scale (QSIS) inhomogeneity using a constant frequency external drive at ion-acoustic time scales, resulting in ion trapped particle instability (ITPI), wave-wave mode coupling interaction and energy cascading. QSIS thus formed is perturbed by applying small amplitude electron acoustic (EA) mode leading to the several key plasma response features. In this Part II, using electrostatic, unbounded, OpenMP Vlasov-Poisson solver i.e VPPM-OMP 1.0, we have investigated the formation of various phase space vortices (PSV) (generated using two step or one step time dependent downward frequency chirping drives) in the presence of background QSIS inhomogeneity obtained in Part I. In addition, we have also performed one to one comparison of individual cases with their homogeneous counterparts with exact simulation parameters. In presence of QSIS inhomogeneity, we have observed interesting phenomenon such as early onset of Langmuir (LAN) mode, suppression of PSV sizes, omission of PSVs when compared to the homogeneous cases. Also, for different two step or one step downward chirp perturbation cases, particle trapping or untrapping fractions and its response to the increasing chirp intervals are respectively reported.   

\end{abstract}
%
\vspace{2pc}
\submitto{\PS}
%
\maketitle
%
%
\section{Introduction}
\label{KIKE_Introduction}

In the Part I of the companion paper [Ref. Part I], we have extensively discussed about the theoretical \cite{bgk1957,oneil1965}, experimental \cite{lynov_1979,Saeki_1979,Danielson_2004} and computational simulation \cite{manfredi1997,raghunathan2013,Saini2018,pallavi2016,pallavi2017,pallavithesis,sanjeev2021,Pandey_2021_TPI_1,Pandey_2021_TPI_2,pandey_2022_KAW,Pandey_2024,sanjeevthesis} investigations reported for the Bernstein-Greene-Kruskal (BGK) modes. In particular, we have briefed about the low amplitude non-linear BGK mode, commonly known as electron acoustic waves (EAW), reported first by Holloway and Dorning in 1991 \cite{Holloway_Dorning_1991}. Several investigations incorporating experimental evidence \cite{Anderegg_2009,Anderegg_PRL_2009} alongwith excitation, stability and associated parametric instabilities \cite{valentini_2006,valentni_2025,Rivera_2025,Berk_1970,DEPACKH_1962,manfredi2000} were also reported. As elaborated in the companion paper Part I, several studies have utilized the time dependent or chirp external drivers to excite various BGK modes in bounded \cite{Breizman_1997,Eremin_2002,Fajans_2003,Friedland_2004,Peinetti_2005} and periodic \cite{pallavi2016,pallavi2017,pallavithesis} plasma systems. It is well known that both the abrupt and adiabatic external electric field drives, with constant frequency and wavenumber alongwith linear amplitude, generates plasma modes with its associated harmonics over a range of frequency values \cite{Anderegg_2009,Schamel_2000,Holloway_Dorning_1991,valentini_2006}. In another theoretical work, Berk et al. \cite{Berk_1997,Berk_Briezman_1997,Eremin_Berk_2002} predicted that spontaneous coherent structures can be generated by upshifting or downshifting of frequency $(\omega)$ near a kinetic instability threshold. Also, in the other works related to Penning-Malmberg trap pure ion experimental plasmas, frequency sweeping study has been performed to excite these BGK modes \cite{Bertsche_2003,Friedland_2004,Peinetti_2005,Anderegg_2009}.

In the recent past, for Maxwellian and Nonextensive $q$ distribution [non-Maxwellian] \cite{pallavi2016,pallavi2017} have demonstrated the existence of a window for chirped external drive frequency, which leads to the formation of giant phase space vortex [PSV]. In this process, when the external drive is chirped, its effective coupling was reported to increase both streaming of ``trapped" and ``untrapped" particle fractions. In Part I companion paper, using OpenMP VPPM-OMP 1.0 \cite{sanjeev2021,Pandey_2021_TPI_1,Pandey_2021_TPI_2,pandey_2022_KAW,Pandey_2024,sanjeevthesis} solver and constant frequency external electric field drive, in a 1D collisionless, electrostatic, unbound plasma system, we have obtained a quasi-stationary ion scale (QSIS) inhomogeneity of scale $k_{eq}/k_{min}=m=2$ [Symbol definitions are given in Part I] where $k_{min}=0.4$. In addition EAW perturbation with frequency $\omega_{EA}^{P}=0.624$ and amplitude $E_{0}^{P}=0.025$ is launched on top of QSIS inhomogeneity to study the evolutionary dynamics of EAW modes which consist of transient phase space vortex (PSV) formation and generation of various frequency associated to EAW and LAN modes [Ref. Part I]. All the numerical experiments in Part I were performed with constant frequency electric field drives. As discussed previously about frequency sweeping process, an obvious next step is to investigate these evolutionary dynamics with time dependent or chirped frequency drives in the presence of background  QSIS inhomogeneity.      

In the present work i.e Part II, we have investigated the long time stability as well as the formation dynamics of various types of phase space vortices (PSV) structures [i.e EAW, LAN, Honeycomb (HC) structures] generated using chirped frequency drive in the presence of background QSIS inhomogeneity. In the present paper, our computational efforts are categorized into two ways in the range $\omega/\omega_{pe}=1.0$ to $\omega/\omega_{pe}=2.0$ mainly, two step frequency chirping and one step frequency chirping method. Details about each of the respective methodologies are elaborated in the subsequent sections. In addition, we have also presented comparative simulation results of each case with homogeneous plasma scenarios with exactly similar simulation parameters. It enables us to understand the effect of background QSIS inhomogeneity on certain kinetic features such as electron phase space dynamics, mode coupling and trapping/untrapping fractional atributes of these cases.

This paper is organized as follows: In Sec. \ref{QSISMS2_Mathematical_model_Numerical_scheme}, we present the model equations including the equations for modified external electric field drive, simulation domain and chirping strategies. Followed by simulation results in Sec. \ref{QSISMS2_Simulation_Results} which consists of two Step Chirping Method [in Sec. \ref{QSISMS2_Two_Step_Method}], Large Phase space Vortex Structure (LPSV) Case [in Sec. \ref{QSISMS2_One_Step_Method_LPSV}], Honeycomb Structure (HC) Case [in Sec. \ref{QSISMS2_One_Step_Method_HC}] and Response of the system to various Chirp intervals [in Sec. \ref{QSISMS2_One_Step_Method_Chirp_Intervals}]. Finally we conclude in Sec. \ref{QSISMS2_Discussion_conclusion}.
 
\section{Mathematical model and Numerical scheme}
\label{QSISMS2_Mathematical_model_Numerical_scheme}

In order to carry out the above said studies, we solve the set of coupled Vlasov-Poisson equation [Eq. (1)-(3) of the companion paper Part I] in the limit of kinetic ions and kinetic electrons using OpenMP VPPM-OMP 1.0 solver. In this work i.e Part II, in the presence of background QSIS inhomogeneity, we are interested to see the response of various types of phase space vortices [PSVs i.e EAW/LAN/Honeycomb like structures] generated using constant frequency and chirped frequency drives. The expression for the total electric field [Eq. (4) of the companion paper Part I] is defined as,
\begin{equation}
E_{T}(x,t)= E_{s}(x,t) + E_{IAW}^{Equil}(x,t) + E_{EAW}^{Pert}(x,t) + E_{Chirp}^{Ext}(x,t)
\label{EQ_1}
\end{equation} 
\begin{equation}
E_{Chirp}^{Ext}(x,t)= E_{c}^{D}sin(k_{c}x \pm \omega_{c}^{D}t)
\label{EQ_2}
\end{equation} 
where $E_{Chirp}^{Ext}(x,t)$ is the external chirped electric field drive applied on top of the created QSIS inhomogeneity obtained in the companion paper Part I, $E_{c}^{D}$ is the amplitude of the time dependent cirpped electric field drive, $k_{c}$ is the scale length, $\omega_{c}$ is the time dependent frequency. Rest of all the quantites have their usual meaning as defined in the companion paper Part I. Note that the chirp term in Eqs. \ref{EQ_1} and \ref{EQ_2} which was not applied in Part I companion paper. Details about the advection and time stepping scheme used in Vlasov-Poisson solver VPPM-OMP 1.0 is elaborated in the companion paper Part I. 

\begin{figure}
\centerline{\includegraphics[scale=0.40]{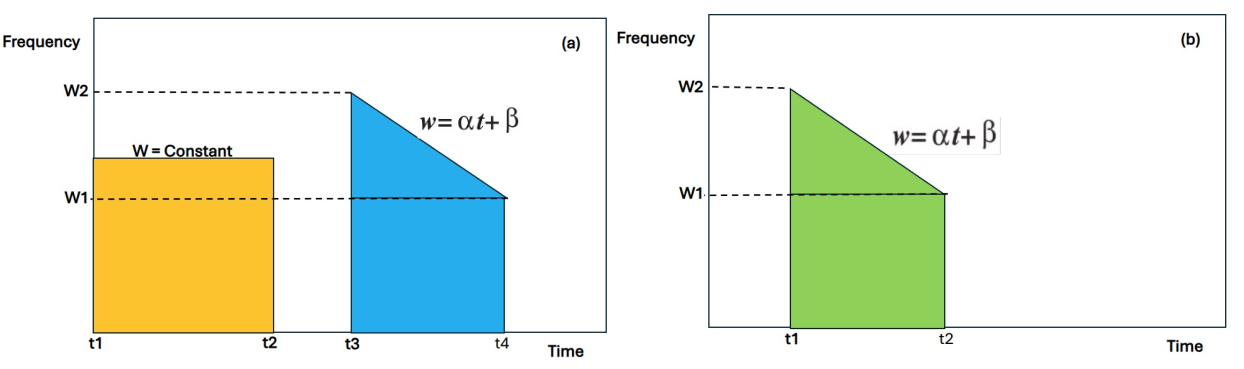}}
\caption{A cartoon figure of frequency versus time i.e $(\omega,t)$ showing frequency turn on-off of an external electric field drive. Each (a) and (b) cases correspond to two step and one step frequency chirping methodologies respectively. In (a) i.e two step chirping process, constant frequency EAW drive $[\omega_{EA}^{P}]$ is applied from $t_{1}<t<t_{2}$ and downward chirped frequency $[\omega=\alpha t + \beta]$ drive is applied from $t_{3}<t<t_{4}$ where $(\alpha, \beta)$ are the chirp coefficents. Note that between time $t_{2}<t<t_{3}$, no external drive is applied as it is a relaxation phase. Meanwhile, in (b) i.e one step chirping process, downward chirped frequency $[\omega=\alpha t + \beta]$ drive is applied directly between $t_{1}<t<t_{2}$. }
\label{1_CHIRP_SCHEME}
\end{figure}

For the study presented in here, we set the simulation domain in the 1D phase space $(x,v)$ as : $D=[0,L_{max}] \times [-v_{e}^{max},v_{e}^{max}]$, where $L_{max}=2\pi / k_{min}=5\pi$ (as $k_{min}=0.4$) is the system size, $v_{e}^{max}=8.0$, grid discretizations $[N_{x} \times N_{v} =1024 \times 6000]$. Periodic boundary conditions are being implemented in both the space and velocity domains. In the present work, we have used two different methodologies of frequency chirping, namely, first, a two-step downward frequency chirp and second, a one step downward frequency chirp, as shown in Fig. \ref{1_CHIRP_SCHEME}. Fig. \ref{1_CHIRP_SCHEME} illustrates a cartoon figure of frequency versus time i.e $(\omega,t)$ indicating frequency turn on-off of an external electric field drive. Cases (a) and (b) correspond to two step and one step frequency chirping methodologies respectively. In (a) i.e two step chirping process, constant frequency EAW drive is applied from $t_{1}<t<t_{2}$ and then a downward chirped frequency $(\omega=\alpha t + \beta)$ drive is applied from $t_{3}<t<t_{4}$ where $(\alpha, \beta)$ are the chirp coefficents. Note that between time $t_{2}<t<t_{3}$, no external drive is applied. In (b) i.e one step chirping process, downward chirped frequency $[\omega=\alpha t + \beta]$ drive is directly applied between $t_{1}<t<t_{2}$ intervals. Rest of the details of numerical diagnostics and simulation constructs are exactly similar as discussed in the companion paper Part I. In the next section, we will elaborate on the simulation results obtained using the two chirp methodologies in the presence of background QSIS inhomogeneity.

\section{Simulation Results}
\label{QSISMS2_Simulation_Results}

\subsection{Two Step Chirping Method : Constant Frequency Drive + Downward Chirped Frequency Drive}
\label{QSISMS2_Two_Step_Method}

\begin{figure}. 
\centerline{\includegraphics[scale=0.68]{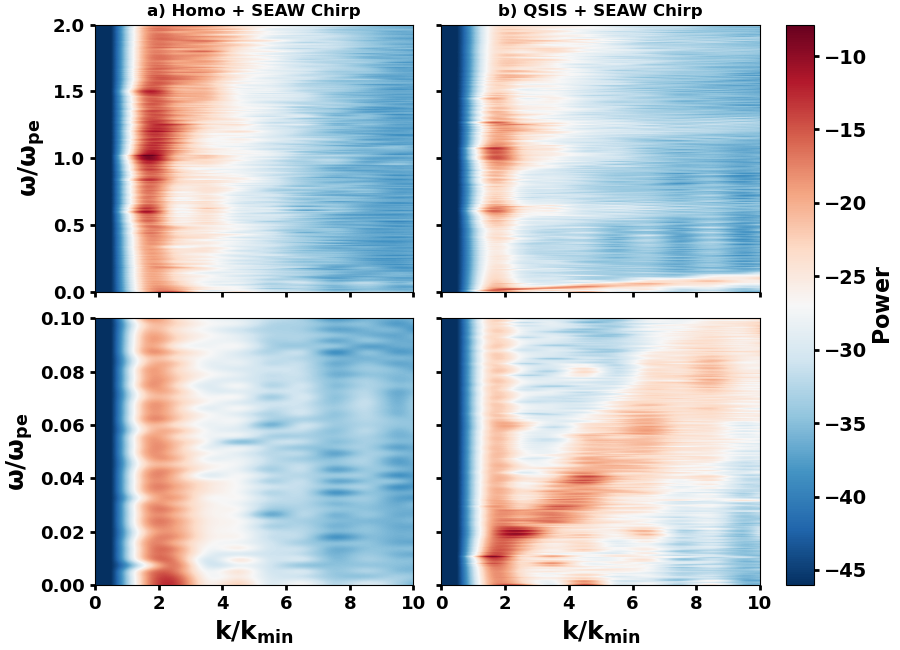}}
\caption{2D $(\omega,k)$ power spectrum plot of (a) Homogeneous + EAW perturbation + Chirp [top and bottom left] and (b) QSIS inhomogeneity + EAW perturbation + Chirp [top and bottom right]  driven plasma cases with perturbation interval $\Delta t_{EA}^{P}=1000 ~\omega_{pe}^{-1}$, $\omega_{EA}^{P}=0.624$, $E_{0}^{P}=0.025$, $k_{p}/k_{min}=1$, $k_{c}/k_{min}=1$, $E_{c}^{D}=0.025$ and chirp interval $\Delta t_{Chirp}=250 ~\omega_{pe}^{-1}$. Comparing (a) and (b), we can observe the supression of higher frequency modes in the range of $\omega / \omega_{pe}=1.2$ to $2.0$ in the QSIS case. Also, one can see the presence and absence of EAW mode, LAN mode and mode coupling signatures in either cases respectively.}
\label{1_2DPS_SEAW}
\end{figure}

In the following two step process, analogous to the work done in Refs. \cite{pallavi2016,pallavi2017}, we drive the system with a constant frequency initially to create a ``seed" flattening in the electron phase space followed by a relaxation period of $\Delta t_{relaxation}=500~\omega_{pe}^{-1}$ ($t_{2}<t<t_{3}$), afterwards, a chirped frequency drive is applied for a period of $\Delta t_{Chirp}=250~\omega_{pe}^{-1}$ which leads to the enhancement of the formed PSVs.In this method, the involved scale lengths of two drives are $k_{p}/k_{min}=1$ and $k_{c}/k_{min}=1$, whereas, the frequency values are $\omega_{EA}^{P}=0.624$ and for the downward chirp part $\omega=\alpha t +\beta$, frequency is swept from $\omega_{c}=1.0$ with chirp coefficents $[\alpha, \beta] : [-5.0 \times 10^{-3},2.0]$. The corresponding time periods of the constant frequency slow electron acoustic (SEAW) perturbation drive, relaxation and chirped frequency drives are : $\Delta t_{EAW}^{\omega=Const}=1000~\omega_{pe}^{-1}$, $\Delta t_{relaxation}=500~\omega_{pe}^{-1}$ and $\Delta t_{Chirp}=250~\omega_{pe}^{-1}$ respectively. We advect the solution till $t=3000~\omega_{pe}^{-1}$ in homogeneous case and till $t=123000~\omega_{pe}^{-1}$ in QSIS inhomogeneity case. The main difference between the work of Ref. \cite{pallavi2016,pallavi2017} and this study is the presence of a non-Maxwellian quasi-stationary ion distribution function as a starting point instead of a Maxwellian ion distribution.


\begin{figure}
\centerline{\includegraphics[scale=0.50]{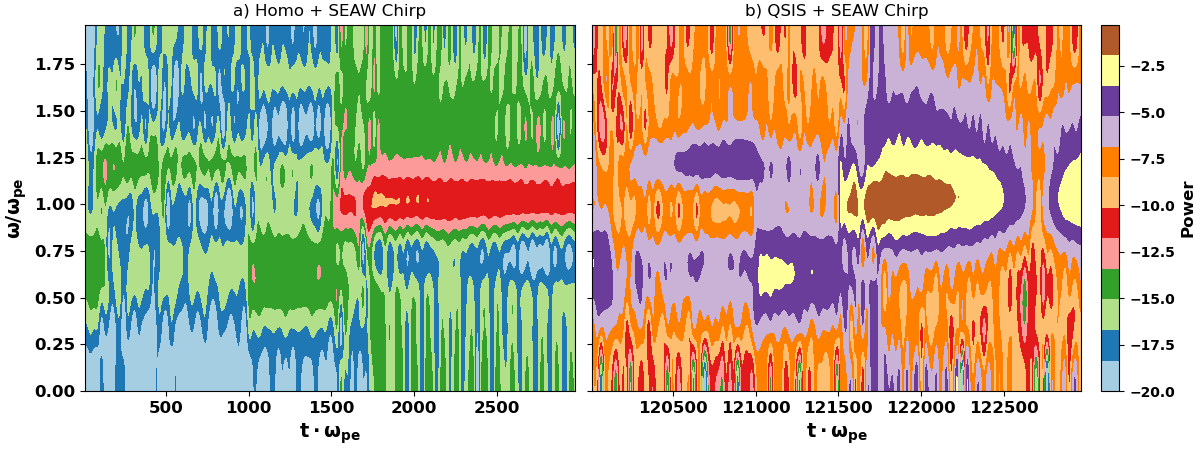}}
\caption{Spectogram plot i.e variation of the generated frequency $(\omega/\omega_{pe})$ with respect to simulation time for (a) Homogeneous $\Delta t_{Chirp}=400~\omega_{pe}^{-1}$+ EAW perturbation + Chirp and (b) QSIS inhomogeneity + EAW perturbation + Chirp driven plasma cases with perturbation interval $\Delta t_{EA}^{P}=1000 ~\omega_{pe}^{-1}$, $\omega_{EA}^{P}=0.624$, $E_{0}^{P}=0.025$, $k_{p}/k_{min}=1$, $k_{c}/k_{min}=1$, $E_{c}^{D}=0.025$ and chirp interval $\Delta t_{Chirp}=250 ~\omega_{pe}^{-1}$ respectively. Both (a) and (b) illustrates the generated frequency signatures during the perturbation, chirp and relaxation periods.}
\label{1_SPECTOGRAM_SEAW}
\end{figure}

In general, when a downward frequency chirp is applied in a homogeneous plasma system, it causes the trapping of particles leading to a giant flatenning of the distribution function alongwith increase in the untrapped particle population. Till the chirp drive is switched on, there is increase in the distribution flatenning. After the drive is switched off, the phase space structures undergoes a weak relaxation leading to a stationary giant non-linear structure embeded with multiple extremas of holes and clumps \cite{pallavi2016,pallavi2017}. Fig. \ref{1_2DPS_SEAW} shows 2D $(\omega,k)$ power spectrum plot of (a) Homogeneous + EAW perturbation + Chirp [top and bottom left] and (b) QSIS inhomogeneity + EAW perturbation + Chirp [top and bottom right] driven plasma cases with perturbation interval $\Delta t_{EA}^{P}=1000 ~\omega_{pe}^{-1}$, $\omega_{EA}^{P}=0.624$, $E_{0}^{P}=0.025$, $k_{p}/k_{min}=1$, $k_{c}/k_{min}=1$, $E_{c}^{D}=0.025$ and chirp interval $\Delta t_{Chirp}=250 ~\omega_{pe}^{-1}$. From Fig. \ref{1_2DPS_SEAW} (a) and (b), we can observe the clear existence of frequency values corresponding to EAW ($\omega_{EAW}=0.624$) and LAN [$\omega_{LAN}=1.284$] modes in both the QSIS inhomogeneity and homogeneous cases. Since, in both the cases, we started the downward chirp from $\omega_{c}=1.0$, so, a considerable fraction of power association is seen around $\omega / \omega_{pe}=1.0, ~ k/k_{min}=1.0$. However, in the zoomed plot which is in the range of 0 to 1.0 i.e bottom left and bottom right plot, we can see the mode coupling signature i.e power distribution among the interacting modes $k/k_{min}=1$ to 6 in the QSIS inhomogeneity case. There is absence of similar phenomenon for the homogeneous case. Also, we have noticed the supression of the generated frequencies in the QSIS inhomogeneity case from the range of $\omega / \omega_{pe}=1.0$ to $2.0$ when compared to the homogeneous case as shown in Fig. \ref{1_2DPS_SEAW} (a) and (b).

Fig. \ref{1_SPECTOGRAM_SEAW} demonstrates spectogram plot i.e variation of the generated frequency $(\omega/\omega_{pe})$ with respect to simulation time for (a) Homogeneous + EAW perturbation + Chirp and (b) QSIS inhomogeneity + EAW perturbation + Chirp driven plasma cases with perturbation interval $\Delta t_{EA}^{P}=1000 ~\omega_{pe}^{-1}$, $\omega_{EA}^{P}=0.624$, $E_{0}^{P}=0.025$, $k_{p}/k_{min}=1$, $k_{c}/k_{min}=1$, $E_{c}^{D}=0.025$ and chirp interval $\Delta t_{Chirp}=250 ~\omega_{pe}^{-1}$ respectively. Both (a) and (b) illustrates the generated frequency signatures during the perturbation, chirp and relaxation periods. As, we started the chirp from $\omega_{c}=1.0$, there is a prominent band around $\omega/\omega_{pe}=1.0$. However, it also demonstrates the nature of frequency generation throughout the temporal evolution in both the cases. The spectrum is more discontinous in QSIS inhomogeneity case compared to the homogeneous case, due to the presence of ion scale background inhomogeneity and wave-wave mode coupling interactions.

\begin{figure}
\centerline{\includegraphics[scale=0.57]{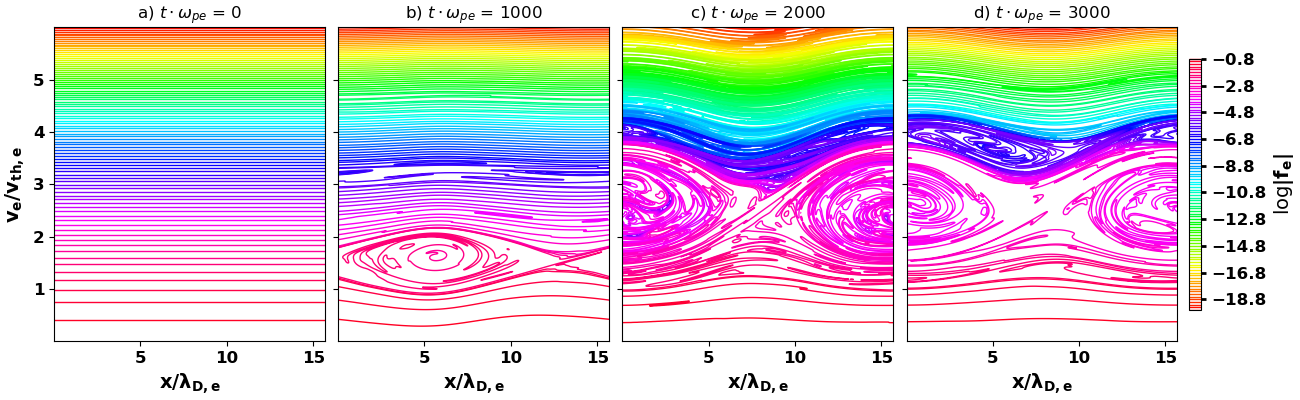}}
\caption{Phase space portrait of electron distribution function $f_{e}(x,v)$ at different times i.e (a) $t=0~\omega_{pe}^{-1}$, (b) $t=1000~\omega_{pe}^{-1}$, (c) $t=2000~\omega_{pe}^{-1}$ and (d) $t=3000~\omega_{pe}^{-1}$ for Homogeneous + EAW perturbation + Chirp  case with perturbation interval $\Delta t_{EA}^{P}=1000 ~\omega_{pe}^{-1}$, $\omega_{EA}^{P}=0.624$, $E_{0}^{P}=0.025$, $k_{p}/k_{min}=1$, $k_{c}/k_{min}=1$, $E_{c}^{D}=0.025$ and chirp interval $\Delta t_{Chirp}=250 ~\omega_{pe}^{-1}$. From (a) to (d), we can observe the formation of EAW and LAN modes in the electron phase space as well as enhanced vortex structures due to downward chirp induced particle trapping.}
\label{1_CP_SEAW_HOMO}
\end{figure}

Fig. \ref{1_CP_SEAW_HOMO} illustrates phase space portrait of electron distribution function $f_{e}(x,v)$ at different times i.e (a) $t=0~\omega_{pe}^{-1}$, (b) $t=1000~\omega_{pe}^{-1}$, (c) $t=2000~\omega_{pe}^{-1}$ and (d) $t=3000~\omega_{pe}^{-1}$ for Homogeneous + EAW perturbation + Chirp  case with perturbation interval $\Delta t_{EA}^{P}=1000 ~\omega_{pe}^{-1}$, $\omega_{EA}^{P}=0.624$, $E_{0}^{P}=0.025$, $k_{p}/k_{min}=1$, $k_{c}/k_{min}=1$, $E_{c}^{D}=0.025$ and chirp interval $\Delta t_{Chirp}=250 ~\omega_{pe}^{-1}$. In Fig. \ref{1_CP_SEAW_HOMO} (b) at the end of $t=1000~\omega_{pe}^{-1}$, we can clearly notice the formation of EAW and LAN modes around corresponding phase velocities $v_{\phi}|_{EAW}=\omega_{EA}^{P}/[k_{p}/k_{min}]=0.624/0.4=1.56$ and $v_{\phi}|_{LAN}=\omega_{LAN}^{P}/[k_{p}/k_{min}]=[\sqrt{1+3(0.4)^{2}}]/0.4=1.2165/0.4=3.0414$ respectively. We have referred this a ``seed" flattening in the distribution. Next, with the application of the downward frequency chirp during $1500~\omega_{pe}^{-1}<t<1750~\omega_{pe}^{-1}$ i.e $\Delta t_{Chirp}=250 ~\omega_{pe}^{-1}$, we have observed the formation of a giant phase space vortex around phase velocity $v_{\phi}=\omega_{c}/[k/k_{min}]=1/0.4=2.50$ alongwith a second hole structure at higher phase velocity $v_{\phi}=4.0$ due to chirp induced particle trapping phenomenon as shown in Fig. \ref{1_CP_SEAW_HOMO} (c). After further relaxation of $1000~\omega_{pe}^{-1}$ at the end of the simulation, steady state vortex is created by combination of both trapped and untrapped particle dynamics during chirp. Apart from the two giant vortices around 2.50 and 4.0, there is a large region of ``separatrices" squashed between these structures. At $t=3000~\omega_{pe}^{-1}$, we have observed a slight particle untrapping around separatrix region alongwith enhancement in the $v_{\phi}=4.0$ vortex structure as shown in Fig. \ref{1_CP_SEAW_HOMO} (d). The obtained steady state is an example of multiple extrema PSV as these structures consists of embedded peaked spikes, holes and clumps.

\begin{figure}
\centerline{\includegraphics[scale=0.57]{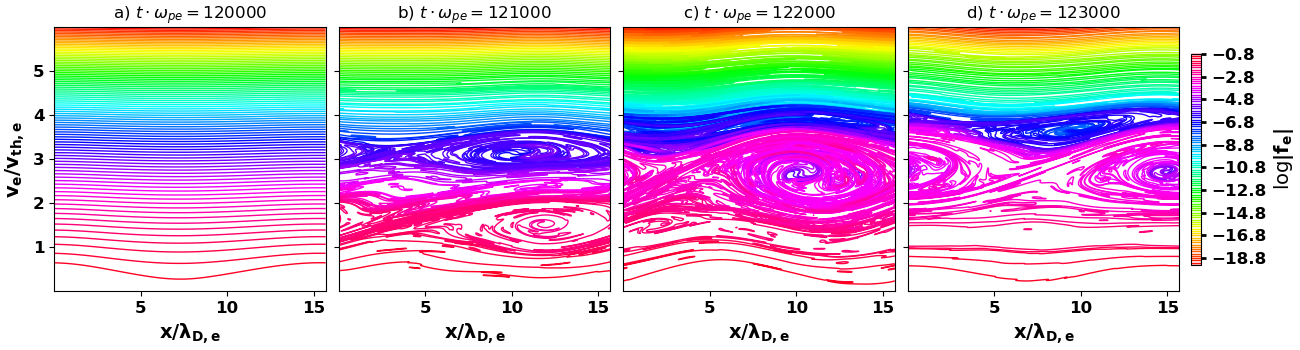}}
\caption{Phase space portrait of electron distribution function $f_{e}(x,v)$ at different times i.e (a) $t=0~\omega_{pe}^{-1}$, (b) $t=1000~\omega_{pe}^{-1}$, (c) $t=2000~\omega_{pe}^{-1}$ and (d) $t=3000~\omega_{pe}^{-1}$ for QSIS inhomogeneity + EAW perturbation + Chirp case with perturbation interval $\Delta t_{EA}^{P}=1000 ~\omega_{pe}^{-1}$, $\omega_{EA}^{P}=0.624$, $E_{0}^{P}=0.025$, $k_{eq}/k_{min}=2$, $k_{p}/k_{min}=1$, $k_{c}/k_{min}=1$, $E_{c}^{D}=0.025$ and chirp interval $\Delta t_{Chirp}=250 ~\omega_{pe}^{-1}$. From (a) to (d), one can observe the formation and enhancement of EAW and LAN mode structures in the electron phase space due to downward chirp induced particle trapping as well as the interaction of these modes with the background ion scale inhomogeneity.}
\label{1_CP_SEAW_QSIS}
\end{figure}

\begin{figure}
\centerline{\includegraphics[scale=0.55]{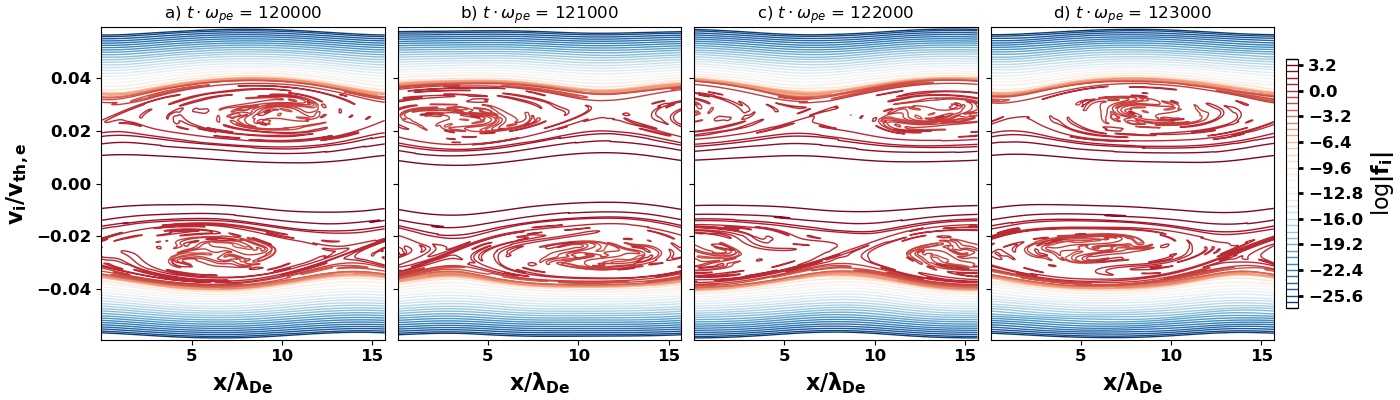}}
\caption{Phase space portrait of ion distribution function $f_{e}(x,v)$ at different times i.e (a) $t=120000~\omega_{pe}^{-1}$, (b) $t=121000~\omega_{pe}^{-1}$, (c) $t=122000~\omega_{pe}^{-1}$ and (d) $t=123000~\omega_{pe}^{-1}$ for QSIS inhomogeneity + EAW perturbation + Chirp case with perturbation interval $\Delta t_{EA}^{P}=1000 ~\omega_{pe}^{-1}$, $\omega_{EA}^{P}=0.624$, $E_{0}^{P}=0.025$, $k_{eq}/k_{min}=2$, $k_{p}/k_{min}=1$, $k_{c}/k_{min}=1$, $E_{c}^{D}=0.025$ and chirp interval $\Delta t_{Chirp}=250 ~\omega_{pe}^{-1}$.}
\label{1_CP_SEAW_QSIS_IONS}
\end{figure}

Fig. \ref{1_CP_SEAW_QSIS} demonstrates phase space portrait of electron distribution function $f_{e}(x,v)$ at different times i.e (a) $t=0~\omega_{pe}^{-1}$, (b) $t=1000~\omega_{pe}^{-1}$, (c) $t=2000~\omega_{pe}^{-1}$ and (d) $t=3000~\omega_{pe}^{-1}$ for QSIS inhomogeneity + EAW perturbation + Chirp case with perturbation interval $\Delta t_{EA}^{P}=1000 ~\omega_{pe}^{-1}$, $\omega_{EA}^{P}=0.624$, $E_{0}^{P}=0.025$, $k_{eq}/k_{min}=2$, $k_{p}/k_{min}=1$, $k_{c}/k_{min}=1$, $E_{c}^{D}=0.025$ and chirp interval $\Delta t_{Chirp}=250 ~\omega_{pe}^{-1}$. Similar to the homogeneous case shown previously, we have observed the enhancement in the phase space vortex structures upon driving the system with a downward chirp time dependent frequency drive as shown in Fig. \ref{1_CP_SEAW_QSIS} (a) to (d). On comparing Fig. \ref{1_CP_SEAW_HOMO} and Fig. \ref{1_CP_SEAW_QSIS}, we arrive at some key differences, first, we have observed the onset and formation of huge PSV embedded with hole and clumps around $v_{\phi}|_{LAN}=3.21$ resonance location at $t=1000 - 121000~\omega_{pe}^{-1}$. It indicates that due to the presence of background QSIS inhomogeneity, particle trapping around $v_{\phi}|_{LAN}$ location enhances even before the EAW perturbation drive ends which is not there in the homogeneous case. Comparing Fig. \ref{1_CP_SEAW_HOMO} [(b) and (c)] and Fig. \ref{1_CP_SEAW_QSIS} [(b) and (c)], we argue that the particle trapping around the separatrix region is more in QSIS inhomogeneity case than the homogeneous case. Secondly, we have seen a reduced size phase space vortex formation at late times after relaxation in the QSIS inhomogeneity case compared to the homogeneous one indicating that the presence of an ion scale inhomogeneity in the background suppresses the vortex formation in the phase space. In order to validate this argument, we will later present the quantitative excess density fraction numbers at various temporal locations. Fig. \ref{1_CP_SEAW_QSIS_IONS} shows Phase space portrait of ion distribution function $f_{e}(x,v)$ at different times i.e (a) $t=120000~\omega_{pe}^{-1}$, (b) $t=121000~\omega_{pe}^{-1}$, (c) $t=122000~\omega_{pe}^{-1}$ and (d) $t=123000~\omega_{pe}^{-1}$ for QSIS inhomogeneity + EAW perturbation + Chirp case with perturbation interval $\Delta t_{EA}^{P}=1000 ~\omega_{pe}^{-1}$, $\omega_{EA}^{P}=0.624$, $E_{0}^{P}=0.025$, $k_{eq}/k_{min}=2$, $k_{p}/k_{min}=1$, $k_{c}/k_{min}=1$, $E_{c}^{D}=0.025$ and chirp interval $\Delta t_{Chirp}=250 ~\omega_{pe}^{-1}$. It indicates that the perturbation and chirp drives applied in the study does not affect the background ion equilibrium.

\begin{figure}
\centerline{\includegraphics[scale=0.57]{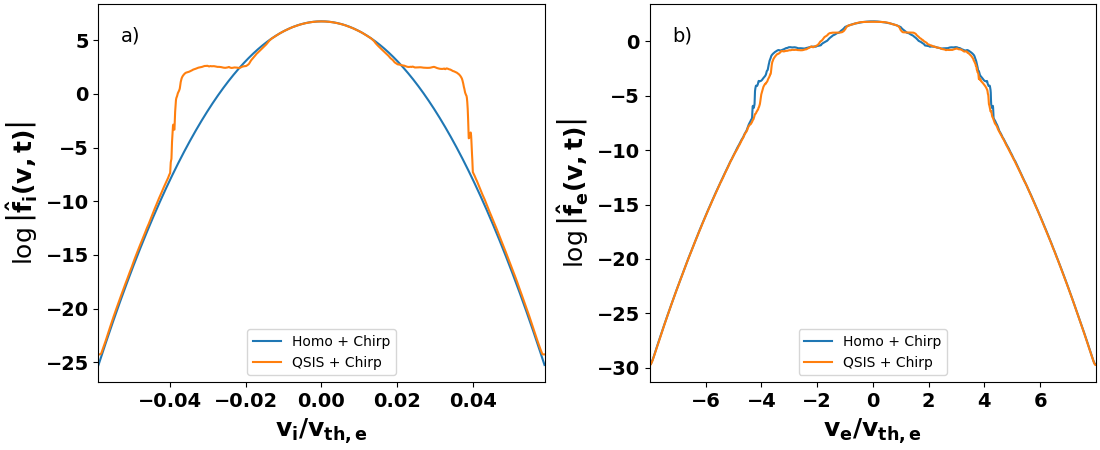}}
\caption{Spatially averaged ion and electron distribution function plots at end times i.e (a) $\hat{f_{i}}(v_{i},t)$ and (b) $\hat{f_{e}}(v_{e},t)$ [defined in Sec. 3 of the companion paper Part I] of both Homogeneous and QSIS inhomogeneity + EAW perturbation + Chirp driven plasma cases with perturbation interval $\Delta t_{EA}^{P}=1000 ~\omega_{pe}^{-1}$, $\omega_{EA}^{P}=0.624$, $E_{0}^{P}=0.025$, $k_{eq}/k_{min}=2$, $k_{p}/k_{min}=1$, $E_{c}^{D}=0.025$, $k_{c}/k_{min}=1$ and chirp interval $\Delta t_{Chirp}=250 ~\omega_{pe}^{-1}$ respectively. Fig (a) and (b) shows the local flattening created in the ion and electron distributions due to wave-particle resonance interactions around respective phase velocities $v_{\phi}^{(i,e)}(\omega_{(i,e)},k)$.}
\label{1_DFI+DFE_SEAW_QSIS+HOMO}
\end{figure}

In Fig. \ref{1_DFI+DFE_SEAW_QSIS+HOMO}, we show spatially averaged ion and electron distribution function plots at end times i.e (a) $\hat{f_{i}}(v_{i},t)$ and (b) $\hat{f_{e}}(v_{e},t)$ [defined in Sec. 3 of the companion paper Part I] of both Homogeneous and QSIS inhomogeneity + EAW perturbation + Chirp driven plasma cases with perturbation interval $\Delta t_{EA}^{P}=1000 ~\omega_{pe}^{-1}$, $\omega_{EA}^{P}=0.624$, $E_{0}^{P}=0.025$, $k_{eq}/k_{min}=2$, $k_{p}/k_{min}=1$, $E_{c}^{D}=0.025$, $k_{c}/k_{min}=1$ and chirp interval $\Delta t_{Chirp}=250 ~\omega_{pe}^{-1}$ respectively. On comparing both QSIS inhomogeneity and homogeneous case spatially averaged ion distributions in Fig. \ref{1_DFI+DFE_SEAW_QSIS+HOMO} (a), it is evident that in the ion phase space there is presence of huge flat top i.e QSIS inhomogeneity structure around $v_{\phi}^{ion}|_{QSIS}=\omega_{IA}^{D}/[k_{eq}/k_{min}]=0.0202/0.8=0.02525$. Meanwhile, ion distribution for homogeneous case remains Maxwellian. Also, comparison of the spatially averaged electron distributions of both the cases in Fig. \ref{1_DFI+DFE_SEAW_QSIS+HOMO} (b), indicates that the flattening or hump created in the distribution function of the homogeneous case (blue one) is larger than the QSIS inhomogeneous case which supports our earlier argument of suppression of phase space vortex formation in the presence of an ion scale inhomogeneous background.

\begin{table}
\caption[ ]{Electron excess density fraction [EDF : $(\delta n/n_{0})$] for homogeneous and $(\delta n_{e}-\delta \hat{n_{e}})/n_{e0}$ for the QSIS inhomogeneity case at $x=L_{max}/8$ estimated at various temporal locations denoted by $\delta_{1}$, $\delta_{2}$ and $\delta_{3}$ respectively (as shown in Fig. \ref{1_EDF_SEAW}), with perturbation interval $\Delta t_{EA}^{P}=1000 ~\omega_{pe}^{-1}$, $\omega_{EA}^{P}=0.624$, $E_{0}^{P}=0.025$, $k_{eq}/k_{min}=2$, $k_{p}/k_{min}=1$, $k_{c}/k_{min}=1$, $E_{c}^{D}=0.025$ and chirp interval $\Delta t_{Chirp}=250 ~\omega_{pe}^{-1}$ . }   
\centering                         
\begin{tabular}{c c c c }           
\hline\hline                        
Case & $\delta_{1}$  & $\delta_{2}$  &  $\delta_{3}$   \\ [1.0ex]    
\hline          
Homogeneous & 0.023 & 0.152 & 0.065  \\          
QSIS Inhomogeneity & 0.035 & 0.118 & 0.026   \\[1ex]
\hline\hline                               
\end{tabular}
\label{TABLE_1}
\end{table}

\begin{figure}
\centerline{\includegraphics[scale=0.045]{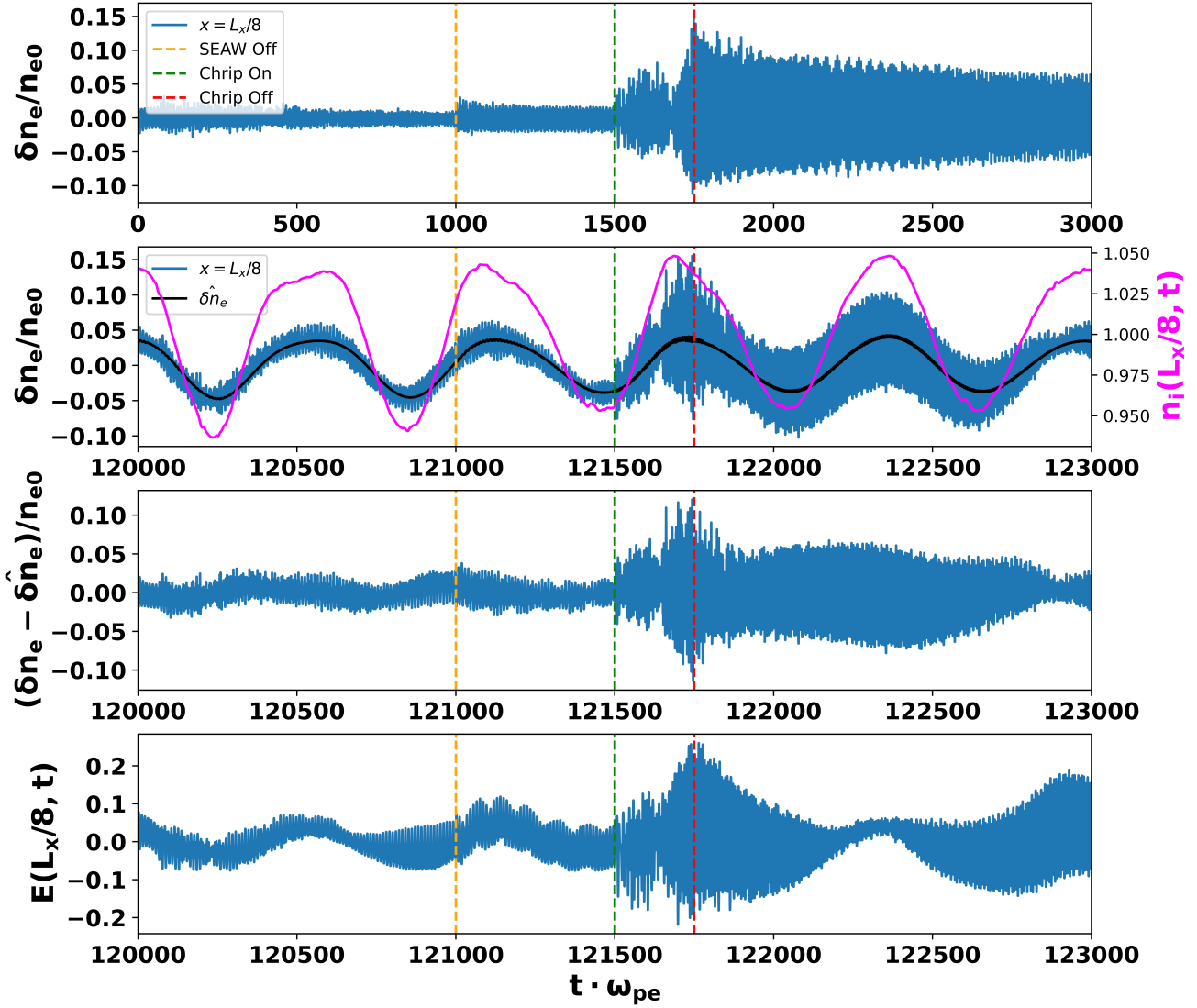}}
\caption{Temporal evolution of (a) homogeneous case excess density fraction (EDF) of electrons i.e $\delta n_{e}/n_{e0}$ at $x=L_{max}/8$, (b) electron EDF at $x=L_{max}/8$, ion density $n_{i}(x=L_{max}/8,t)$ and moving average variation $\delta \hat{n_{e}}$, (c) difference of electron EDF and moving average values i.e $(\delta n_{e} - \delta \hat{n_{e}})/n_{e0}$ at $x=L_{max}/8$, (d) electric field variation at $x=L_{max}/8$ for both Homogeneous and QSIS inhomogeneity + EAW perturbation + Chirp driven plasma cases with perturbation interval $\Delta t_{EA}^{P}=1000 ~\omega_{pe}^{-1}$, $\omega_{EA}^{P}=0.624$, $E_{0}^{P}=0.025$, $k_{eq}/k_{min}=2$, $k_{p}/k_{min}=1$, $E_{c}^{D}=0.025$, $k_{c}/k_{min}=1$ and chirp interval $\Delta t_{Chirp}=250 ~\omega_{pe}^{-1}$ respectively. In the legend, the dotted lines denotes the different switch on/off times of SEAW and chirp drives as denoted by $\delta_{1},~\delta_{2},~\delta_{3}$ in Table \ref{TABLE_1}. }
\label{1_EDF_SEAW}
\end{figure}

In order to quantify our argument, in Table \ref{TABLE_1}, we have tabulated the electron excess density fraction [EDF : $\delta n_{e}/n_{e0}$, defined in the companion paper Part I] for the homogeneous case alongwith the difference of electron EDF and moving average $\delta \hat{n_{e}}$ value i.e $(\delta n_{e}-\delta \hat{n_{e}})/n_{e0}$ for the QSIS inhomogeneity case at various temporal locations denoted by $\delta_{1}$, $\delta_{2}$ and $\delta_{3}$ respectively, where, $\delta_{1}$ is the end of EAW perturbation drive [$t=1000~ or ~121000~\omega_{pe}^{-1}$], $\delta_{2}$ is the end of downward chirped frequency drive [$t=1750~ or ~121750~\omega_{pe}^{-1}$], $\delta_{3}$ is the end of the simulation time [$t=3000~ or ~123000~\omega_{pe}^{-1}$] for respective cases. Fig \ref{1_EDF_SEAW} illustrates the temporal variation of (a) homogeneous case excess density fraction (EDF) of electrons i.e $\delta n_{e}/n_{e0}$, (b) electron EDF, ion density $n_{i}(x=L_{max}/8,t)$ and moving average variation $\delta \hat{n_{e}}$, (c) difference of electron EDF and moving average values i.e $(\delta n_{e} - \delta \hat{n_{e}})/n_{e0}$, (d) electric field variation at $x=L_{max}/8$ for both homogeneous and QSIS inhomogeneity + EAW perturbation + Chirp driven plasma cases with perturbation interval $\Delta t_{EA}^{P}=1000 ~\omega_{pe}^{-1}$, $\omega_{EA}^{P}=0.624$, $E_{0}^{P}=0.025$, $k_{eq}/k_{min}=2$, $k_{p}/k_{min}=1$, $E_{c}^{D}=0.025$, $k_{c}/k_{min}=1$ and chirp interval $\Delta t_{Chirp}=250 ~\omega_{pe}^{-1}$ respectively. In the QSIS inhomogeneous cases, the primary reason behind calculating $(\delta n_{e}-\delta \hat{n_{e}})/n_{e0}$ value is the presence of large background ion density variation at $x=L_{max}/8$ [See Fig. \ref{1_EDF_SEAW} (b)] which influences the electron density patterns leading to a sinusoidal variation  as can be seen in Fig. \ref{1_EDF_SEAW} (b), hence, estimation of density fraction becomes difficult in this case. So, for the QSIS inhomogeneous case, we have calculated $(\delta n_{e}-\delta \hat{n_{e}})/n_{e0}$ using moving average variation $\delta \hat{n_{e}}$ [See Fig. \ref{1_EDF_SEAW}] to get the accurate estimate of trapping/untrapping fractions. Same diagnostic is also implemented throughout in the later QSIS cases of the present work.

\begin{figure}
\centerline{\includegraphics[scale=0.40]{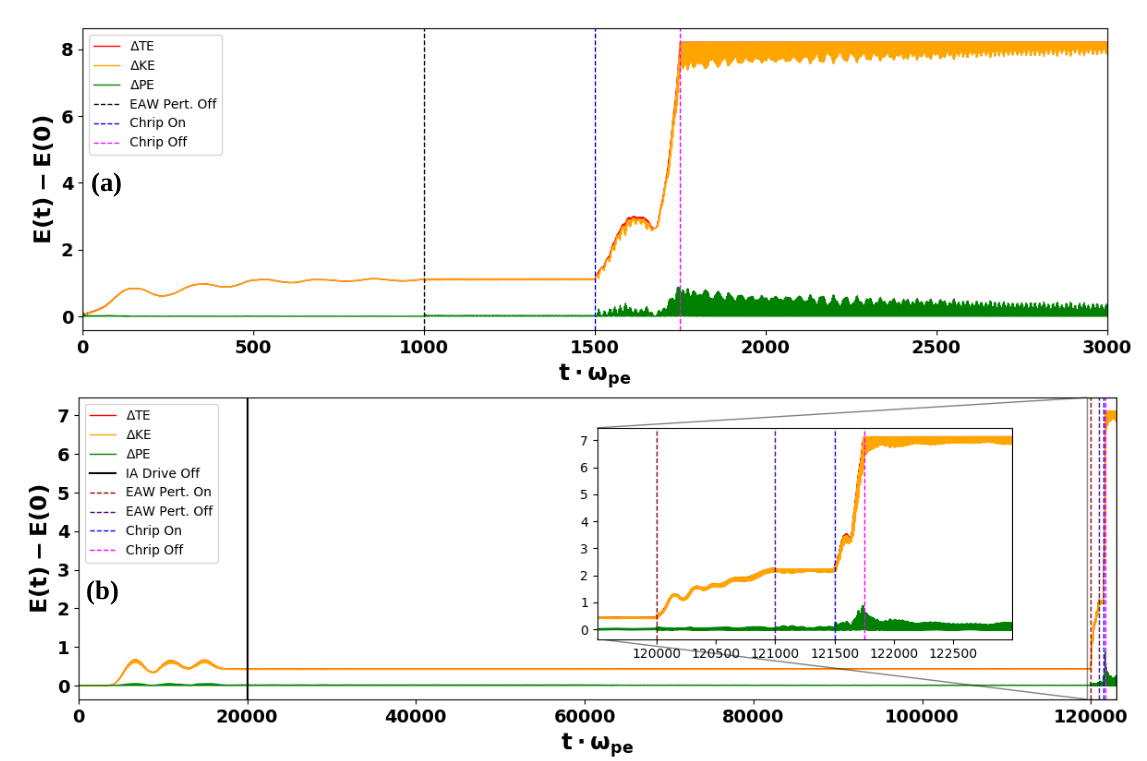}}
\caption{Relative total, potential and kinetic energies $(\Delta TE,~ \Delta PE,~ \Delta KE )$ [defined in Sec. 3 of the companion paper Part I] signatures with respect to time for (a) Homogeneous + EAW perturbation + Chirp and (b) QSIS inhomogeneity + EAW perturbation + Chirp driven plasma cases with perturbation interval $\Delta t_{EA}^{P}=1000 ~\omega_{pe}^{-1}$, $\omega_{EA}^{P}=0.624$, $E_{0}^{P}=0.025$, $k_{eq}/k_{min}=2$, $k_{p}/k_{min}=1$, $k_{c}/k_{min}=1$, $E_{c}^{D}=0.025$, chirp interval $\Delta t_{Chirp}=250 ~\omega_{pe}^{-1}$ and grid resolution $[N_{x} \times N_{v}=1024 \times 6000]$. Solid line denotes the drive switch on and switch off times. Inset plot in (b) shows the zoomed variation of the relative changes in the energies for the QSIS inhomogeneity case.}
\label{1_EC_SEAW_QSIS+HOMO}
\end{figure}

As evident from Fig. \ref{1_EDF_SEAW} (a), for homogeneous case with chirp interval $\Delta t_{Chirp}=250 ~\omega_{pe}^{-1}$, the electron EDF increases rapidly in the interval $1500~\omega_{pe}^{-1} <t < 1750~\omega_{pe}^{-1}$ indicating particle trapping during the chirping drive. After the drive is switched off i.e $t> 1750~\omega_{pe}^{-1}$, the electron EDF relaxes to a finite non-zero value till the end of the simulation. If we refer Table \ref{TABLE_1}, we can clearly see the EDF amplitude variation throughout $\delta_{1}$, $\delta_{2}$ and $\delta_{3}$ locations respectively, indicating, that the particle trapping was at its peak just after the chirp drive was switched off at $t=1750~\omega_{pe}^{-1}$. Meanwhile, for the QSIS inhomogeneity case, analogous trend can be seen from Fig. \ref{1_EDF_SEAW} (c) i.e increase in the $(\delta n_{e}-\delta \hat{n_{e}})/n_{e0}$ value during the applied chirp drive $121500~\omega_{pe}^{-1} <t < 121750~\omega_{pe}^{-1}$. Also, we can observe the relaxation in this estimate at $t=123000~\omega_{pe}^{-1}$ compared to the homogeneous case i.e EDF$|_{\delta_{3}}^{Homo}=0.065$ and EDF$|_{\delta_{3}}^{QSIS}=0.026$ supporting our earlier argument of suppression of PSV formation in the electron phase space due to enhanced late time particle untrapping in the pressence of background QSIS inhomogeneity. In addition, we can also observe the temporal electric field variation around $x=L_{max}/8$ for the QSIS inhomogeneity case showing the rapid increase in the field amplitude during downward chirp drive $121500~\omega_{pe}^{-1} <t < 121750~\omega_{pe}^{-1}$ as shown in Fig. \ref{1_EDF_SEAW} (d). Finite non-zero electric field amplitude at $\delta_{3}=123000~\omega_{pe}^{-1}$ confirms that the obtained late time solution in this case is also a BGK mode.

Fig. \ref{1_EC_SEAW_QSIS+HOMO} shows relative total, potential and kinetic energies $(\Delta TE,~ \Delta PE,~ \Delta KE )$ signatures with respect to time for (a) Homogeneous + EAW perturbation + Chirp and (b) QSIS inhomogeneity + EAW perturbation + Chirp driven plasma cases with perturbation interval $\Delta t_{EA}^{P}=1000 ~\omega_{pe}^{-1}$, $\omega_{EA}^{P}=0.624$, $E_{0}^{P}=0.025$, $k_{eq}/k_{min}=2$, $k_{p}/k_{min}=1$, $k_{c}/k_{min}=1$, $E_{c}^{D}=0.025$, chirp interval $\Delta t_{Chirp}=250 ~\omega_{pe}^{-1}$ and grid resolution $[N_{x} \times N_{v}=1024 \times 6000]$. Solid line denotes the drive switch on and switch off times. Inset plot in (b) shows the zoomed variation of the relative changes in the energies for the QSIS inhomogeneity case. Definitions of all the relative energies are already defined in Sec. 3 of the companion paper Part I. Fig \ref{1_EC_SEAW_QSIS+HOMO} (b) and Fig. 11 of the companion paper Part I are similar from $t=0~\omega_{pe}^{-1}$ to $t=120000~\omega_{pe}^{-1}$, as we created the QSIS background inhomogeneity described in Part I. 

From Fig \ref{1_EC_SEAW_QSIS+HOMO} (a), we can see that during the EAW perturbation interval $0~\omega_{pe}^{-1}<t<1000~\omega_{pe}^{-1}$, only significant changes are registered in the $\Delta TE$ and $\Delta KE $ signatures. Also, $(\Delta TE,~ \Delta PE,~ \Delta KE )$ signatures remains constant in the duration $1000~\omega_{pe}^{-1}<t<1500~\omega_{pe}^{-1}$. But after the downward chirp drive is applied $1500~\omega_{pe}^{-1}<t<17500~\omega_{pe}^{-1}$ i.e chirp interval $\Delta t_{Chirp}=250 ~\omega_{pe}^{-1}$, there is a surge in the $\Delta TE$ and $\Delta KE $ signatures as well as relative $\Delta PE$ values due to increased chirp induced particle trapping. Afterwards, beyond $t>1750~\omega_{pe}^{-1}$, the relative energy signatres saturates to a finite non-zero value indicating good energy conservation and stable solution with grid resolution $[N_{x} \times N_{v}=1024 \times 6000]$. Also, from \ref{1_EC_SEAW_QSIS+HOMO} (b) inset plot, we can clearly observe that the change in the relative difference of the energies in QSIS inhomogeneity case during $120000~\omega_{pe}^{-1}<t<123000~\omega_{pe}^{-1}$ is analogous to the homogeneous case [Fig. \ref{1_EC_SEAW_QSIS+HOMO} (a)]. However, the saturation of the $\Delta TE$ and $\Delta KE $ signatures occurs at a lower value in the QSIS inhomogeneity case compared to the homogeneous case. In the subsequent sections, we will present the simulation results associated with one step chirping methodology generating large phase space vortices (LPSV) and honeycomb (HC) structures respectively. 

\subsection{One Step Chirping Method}
\label{QSISMS2_One_Step_Method}

\begin{figure}
\centerline{\includegraphics[scale=0.68]{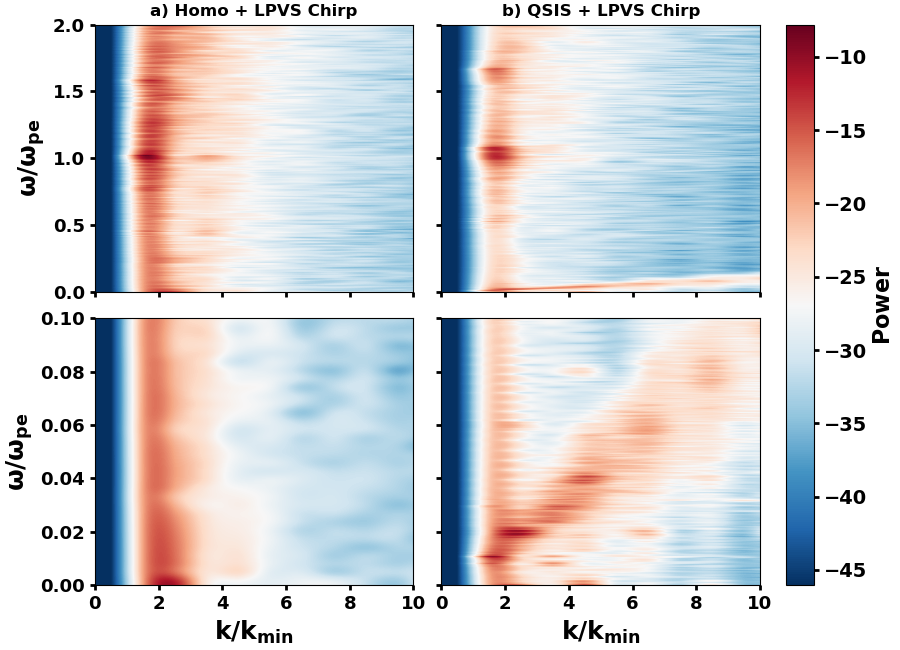}}
\caption{2D $(\omega,k)$ power spectrum plot of (a) Homogeneous + LPSV Chrip [top and bottom left] and (b) QSIS inhomogeneity + LPSV chirp [top and bottom right] driven plasma cases with inhomogeneity scale $k_{eq}/k_{min}=2$, $k_{c}/k_{min}=1$, $E_{c}^{D}=0.025$ and chirp interval $\Delta t_{Chirp}=250 ~\omega_{pe}^{-1}$. Comparing (a) and (b), we can clearly observe the supression of higher frequency in the range of $\omega / \omega_{pe}=0.8$ to $2.0$ in the QSIS case alongwith the difference in the mode coupling signatures of either cases respectively.}
\label{1_2DPS_LPSV}
\end{figure}

In this section, we will implement one step chirping process of two different types, i.e, one for the large phase space vortices (LPSV) and second, for the honeycomb (HC) like transient phase space structures. Similar methodology was implemented by Author \cite{pallavi2016,pallavi2017} in their previous work. Additionally, we will also discuss the response of the system to different chirp intervals alongwith its comparison with the homogeneous case counterparts.

\subsubsection{Large Phase space Vortex Structure (LPSV) Case : }
\label{QSISMS2_One_Step_Method_LPSV}

Numerical experiments were performed with the downward chirped frequency drive which is applied to the plasma right at $t=0~\omega_{pe}^{-1}$ for homogeneous case and at $t=120000~\omega_{pe}^{-1}$ for QSIS inhomogeneity case. The drive is applied for a time interval of $\Delta t_{Chirp}=250~\omega_{pe}^{-1}$ where frequency is swept downwards from $\omega=1.0$ without ``seed" flatenning as discussed in the previous case Sec. \ref{QSISMS2_Two_Step_Method}. Also, the system relaxes for an interval of $\Delta t_{relaxation}=1250~\omega_{pe}^{-1}$. As reported by Authors \cite{pallavi2016,pallavi2017}, we have also found out in these cases that with background QSIS inhomogeneity, the initial ``seed" flatenning is not a crucial component but the chirp is and the PSV structures formed by direct chirping sustains as steady state structures long after the drive is turned off. These cases further demonstrates that chirping can be performed in steps. 

As seen in the previous studies \cite{pallavi2016,pallavi2017}, generally, despite small chirp drive amplitude $E_{c}^{D}$ [much below ``linear limit"], it causes increased particle trapping and simultaneously increase in the kinetic energy which in turn promote the formation of LPSV with tremendous structural complexity in the phase space that sustains till the end of the simulation run. Typically, one observes huge flatenning in the velocity distribution function which grows till the drive is switched on. Weak relaxation alongwith a giant flat region is seen, when the chirp drive is switched off. So, we were curious to investigate these key features of PSV formation in the presence of background QSIS inhomogeneity. Following are the comparative observations we obtained from simulations results performed with exact parameter sets for both the homogeneous and QSIS inhomogeneous cases respectively.

\begin{figure}
\centerline{\includegraphics[scale=0.50]{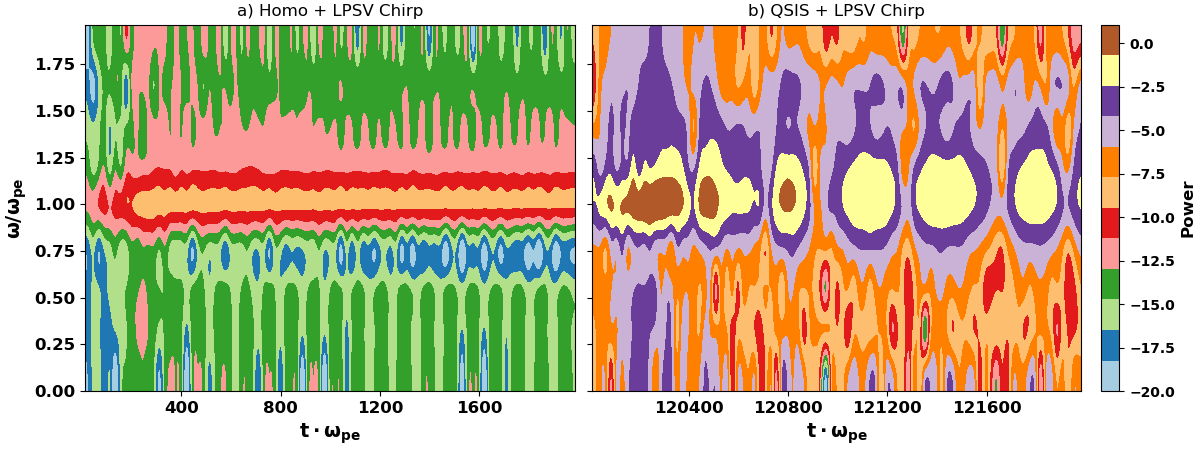}}
\caption{Spectogram plot for (a) Homogeneous + LPSV Chirp and (b) QSIS inhomogeneity + LPSV Chirp driven plasma cases with inhomogeneity scale $k_{eq}/k_{min}=2$, $k_{c}/k_{min}=1$, $E_{c}^{D}=0.025$ and chirp interval $\Delta t_{Chirp}=250 ~\omega_{pe}^{-1}$. From (a) and (b), we can see the continous and discrete nature of frequency generation during the perturbation, chirp and relaxation periods in homogeneous and QSIS inhomogeneity cases respectively.}
\label{1_SPECTOGRAM_LPSV}
\end{figure}

Fig. \ref{1_2DPS_LPSV} illustrates $(\omega,k)$ power spectrum plot of (a) Homogeneous + LPSV Chrip [top and bottom left] and (b) QSIS inhomogeneity + LPSV chirp [top and bottom right]  driven plasma cases with inhomogeneity scale $k_{eq}/k_{min}=2$, $k_{c}/k_{min}=1$, $E_{c}^{D}=0.025$ and chirp interval $\Delta t_{Chirp}=250 ~\omega_{pe}^{-1}$. On comparing Fig. \ref{1_2DPS_LPSV} (a) and (b), we can say that, since, the chirp is applied from $\omega_{c}=1.0$, so, in both the cases we can see the accumulation of power around $\omega/\omega_{pe}=1.0$ which corresponds to $k/k_{min}=1.0$. Alongwith it, we have observed power distribution around $\omega/\omega_{pe}=1.28$ which corresponds to the LAN modes in both the cases as shown in Fig. \ref{1_2DPS_LPSV} (a) and (b) [top left and top right]. Additionally, due to wave-wave mode coupling interaction phenomenon in the QSIS inhomogeneous case, we have observed power distribution among interacting sidebands at very low frequency values in the range of $\omega/\omega_{pe}=0$ to $\omega/\omega_{pe}=1$ [Fig. \ref{1_2DPS_LPSV} (bottom right)]. Meanwhile, there is an absence of such power distribution for homogeneous case as shown in Fig. \ref{1_2DPS_LPSV} (bottom left). Also, on comparing Fig. \ref{1_2DPS_LPSV} [(a) and (b)], we have observed that the power spectrum between the range of $\omega/\omega_{pe}=0.75$ to $\omega/\omega_{pe}=2.0$ is more descrete in the QSIS inhomogeneous case than the homogeneous chirp driven case which is also seen the spectogram diagnostic plot as shown in Fig. \ref{1_SPECTOGRAM_LPSV}.

Fig. \ref{1_SPECTOGRAM_LPSV} shows the spectogram plot for (a) Homogeneous + LPSV Chirp and (b) QSIS inhomogeneity + LPSV Chirp driven plasma cases with inhomogeneity scale $k_{eq}/k_{min}=2$ and chirp interval $\Delta t_{Chirp}=250 ~\omega_{pe}^{-1}$. Comparison of both Fig. \ref{1_SPECTOGRAM_LPSV} (a) and (b), clearly indicates the discontinuity of frequency band around $\omega_{c}=1.0$ in the QSIS inhomogeneous case, whereas, it is continuos till the end of the simulation in the homogeneous case. We suspect this discontinuity in the frequency signature is primarly due to the presence of ion scale inhomogeneity. In addition, generation of high amplitude temporal frequency regions are seen in the QSIS case compared to the homogeneous case when the chirp drive is switched off after a chirp interval of $\Delta t_{Chirp}=250~\omega_{pe}^{-1}$.

\begin{figure}
\centerline{\includegraphics[scale=0.57]{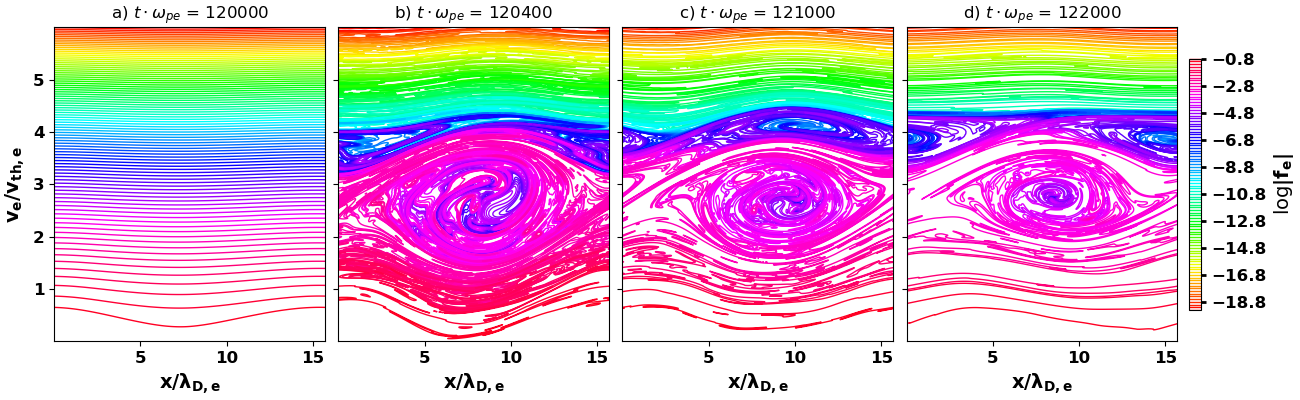}}
\caption{Phase space portrait of electron distribution function $f_{e}(x,v)$ at different times i.e (a) $t=0~\omega_{pe}^{-1}$, (b) $t=120400~\omega_{pe}^{-1}$, (c) $t=121000~\omega_{pe}^{-1}$ and (d) $t=122000~\omega_{pe}^{-1}$ for QSIS inhomogeneity + LPSV Chirp case with inhomogeneity scale $k_{eq}/k_{min}=2$,$k_{c}/k_{min}=1$, $E_{c}^{D}=0.025$ and chirp interval $\Delta t_{Chirp}=250 ~\omega_{pe}^{-1}$. one can observe the formation and enhancement of the large PSV structures in the electron phase space due to downward chirp induced particle trapping alongwith the interaction of these modes with the background ion scale inhomogeneity.}
\label{1_CP_LPSV_QSIS_DELT_250}
\end{figure}

In Fig. \ref{1_CP_LPSV_QSIS_DELT_250}, we demonstrate the phase space portrait of electron distribution function $f_{e}(x,v)$ at different times i.e (a) $t=0~\omega_{pe}^{-1}$, (b) $t=120400~\omega_{pe}^{-1}$, (c) $t=121000~\omega_{pe}^{-1}$ and (d) $t=122000~\omega_{pe}^{-1}$ for QSIS inhomogeneity + LPSV Chirp case with inhomogeneity scale $k_{eq}/k_{min}=2$ and chirp interval $\Delta t_{Chirp}=250 ~\omega_{pe}^{-1}$. At  $t=120000~\omega_{pe}^{-1}$, when the downward frequency chirp is applied from $\omega_{c}=1.0$ [i.e in phase velocity range $v_{\phi}^{c}=\omega_{c}/k=1.0/0.4=2.5$], a ``dip" forms at electron velocity $v_{\phi}^{c}=2.5$ with small amount of trapped particles in it, consequently, as the frequency decreases, there is growth in the size of the ``dip" with addition of more and more particles till the chirp drive is switched on. As one observe from Fig. \ref{1_CP_LPSV_QSIS_DELT_250} (b), (c) and (d), there is a significant density of trapped particles forming a giant electron hole around $v_{\phi}^{c}=2.5$ and at higher phase velocity $v_{\phi}^{c}=4.0$ (approx). After the chirp drive $E_{Chirp}^{Ext}(x,t)$ is switched off at $t=120250~\omega_{pe}^{-1}$, we let the system undergo relaxation process resulting into stationary hole structures that contains peaked spikes and embeded holes in it along with a ``shark"- like structure i.e group of particles moving together within the LPSV. Similar observations were reported in Ref. \cite{pallavi2016,pallavi2017} for homogeneous Vlasov-Poisson plasma cases. Also, some degree of particle untrapping is evident till the end of simulation due to the relaxation as shown in Fig. \ref{1_CP_LPSV_QSIS_DELT_250} (d). We will elaborate more on these particle trapping and untrapping in the next section [Sec. \ref{QSISMS2_One_Step_Method_Chirp_Intervals}] using excess density fraction diagnostics defined in the companion paper Part I.

\begin{figure}
\centerline{\includegraphics[scale=0.55]{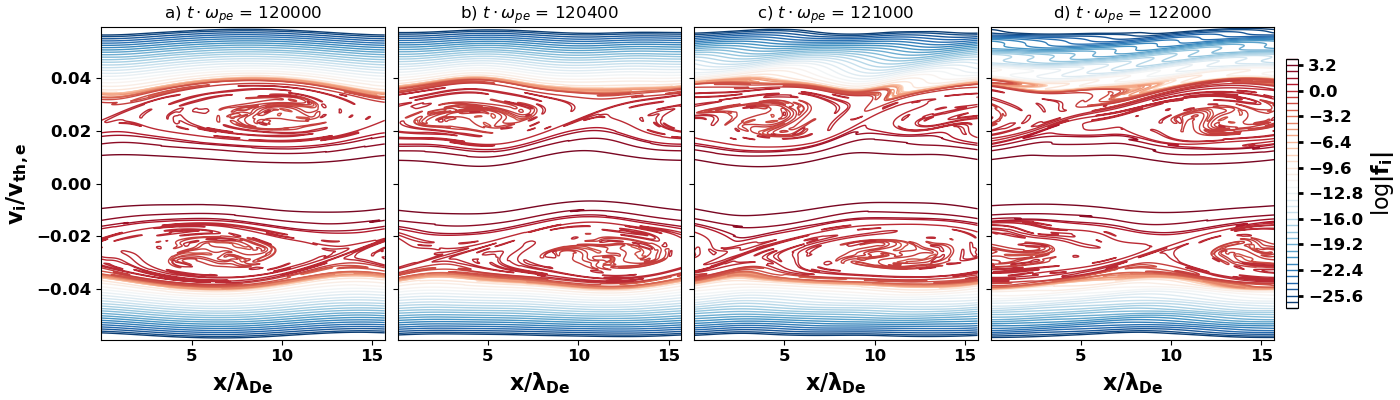}}
\caption{Phase space portrait of ion distribution function $f_{i}(x,v)$ at different times i.e (a) $t=0~\omega_{pe}^{-1}$, (b) $t=120400~\omega_{pe}^{-1}$, (c) $t=121000~\omega_{pe}^{-1}$ and (d) $t=122000~\omega_{pe}^{-1}$ for QSIS inhomogeneity + LPSV Chirp case with inhomogeneity scale $k_{eq}/k_{min}=2$, $k_{c}/k_{min}=1$, $E_{c}^{D}=0.025$ and chirp interval $\Delta t_{Chirp}=250 ~\omega_{pe}^{-1}$. one can observe the formation and enhancement of the large PSV structures in the electron phase space due to downward chirp induced particle trapping alongwith the interaction of these modes with the background ion scale inhomogeneity.}
\label{1_CP_LPSV_QSIS_DELT_250_ION}
\end{figure}

Fig. \ref{1_CP_LPSV_QSIS_DELT_250_ION} shows phase space portrait of ion distribution function $f_{i}(x,v)$ at different times i.e (a) $t=0~\omega_{pe}^{-1}$, (b) $t=120400~\omega_{pe}^{-1}$, (c) $t=121000~\omega_{pe}^{-1}$ and (d) $t=122000~\omega_{pe}^{-1}$ for QSIS inhomogeneity + LPSV Chirp case with inhomogeneity scale $k_{eq}/k_{min}=2$, $k_{c}/k_{min}=1$, $E_{c}^{D}=0.025$ and chirp interval $\Delta t_{Chirp}=250 ~\omega_{pe}^{-1}$. Fig. \ref{1_CP_LPSV_QSIS_DELT_250_ION} (a) to (d) demonstrates that there is no significant effect of electron scale perturbation drive and chirp frequency drives on the created steady state non-stationary QSIS inhomogeneity in the ion phase space. Both the electron and ion scale modes co-exists together in their respective phase spaces.

\begin{figure}
\centerline{\includegraphics[scale=0.57]{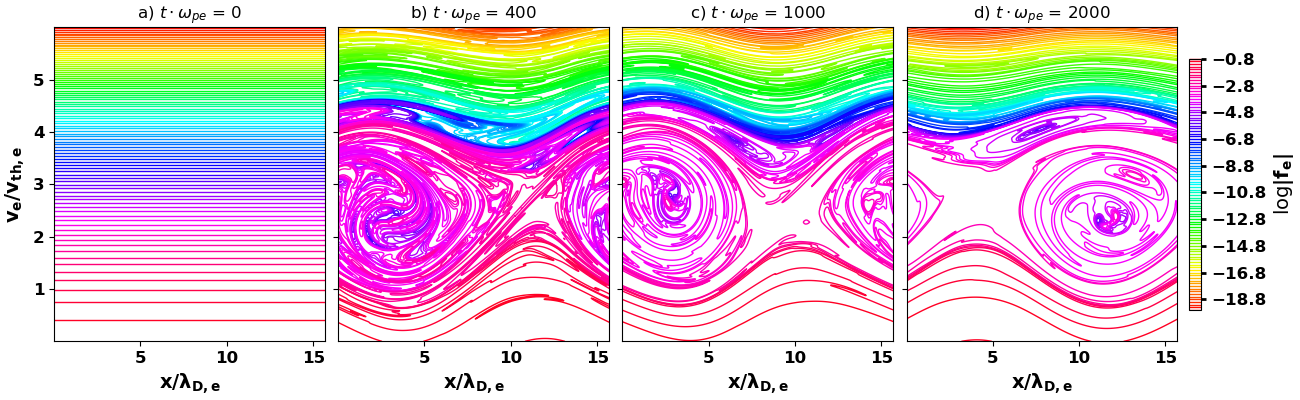}}
\caption{Phase space portrait of electron distribution function $f_{e}(x,v)$ at different times i.e (a) $t=0~\omega_{pe}^{-1}$, (b) $t=400~\omega_{pe}^{-1}$, (c) $t=1000~\omega_{pe}^{-1}$ and (d) $t=2000~\omega_{pe}^{-1}$ for Homogeneous + LPSV Chirp case with $k_{c}/k_{min}=1$, $E_{c}^{D}=0.025$ chirp interval $\Delta t_{Chirp}=250 ~\omega_{pe}^{-1}$. Fig (a) to (d) shows the complete picture of particle trapping enhancement due to downward frequency chirp and its late time relaxation.}
\label{1_CP_LPSV_HOMO_DELT_250}
\end{figure}

In Fig. \ref{1_CP_LPSV_HOMO_DELT_250}, we demonstrate phase space portrait of electron distribution function $f_{e}(x,v)$ at different times i.e (a) $t=0~\omega_{pe}^{-1}$, (b) $t=400~\omega_{pe}^{-1}$, (c) $t=1000~\omega_{pe}^{-1}$ and (d) $t=2000~\omega_{pe}^{-1}$ for Homogeneous + LPSV Chirp case with $k_{c}/k_{min}=1$, $E_{c}^{D}=0.025$ chirp interval $\Delta t_{Chirp}=250 ~\omega_{pe}^{-1}$. Fig (a) to (d) shows the complete picture of particle trapping enhancement due to downward frequency chirp and its late time relaxation. At $t=0~\omega_{pe}^{-1}$, exactly similar downward frequency chirp is applied with the starting chirping frequency $\omega_{c}=1.0$ for comparative investigations. It creates a complex phase space vortex around $v_{\phi}=2.5$ and $v_{\phi}=4.0$ with peaked spikes, embeded holes, and ``Shark" like LPSV structure for homogeneous case. Alternatively, we can observe, particle untrapping during relaxation after the chirp drive is switched off as shown in Fig. \ref{1_CP_LPSV_HOMO_DELT_250} (d). Comparing Fig. \ref{1_CP_LPSV_QSIS_DELT_250} (b) and Fig. \ref{1_CP_LPSV_HOMO_DELT_250} (b) i.e $f_{e}(x,v,t=400~-~120400~\omega_{pe}^{-1})$, we can conclude that there is more fraction of particle trapping in the separatix region of LPSV structure formed in the presence of background QSIS inhomogeneity case compared to the homogeneous case. Also, it is clearly evident that the untrapping fraction of particles during relaxation period [$120250~\omega_{pe}^{-1}<t< 122000~\omega_{pe}^{-1}$] is higher in homogeneous case compared to QSIS inhomogeneous case as shown in Fig. \ref{1_CP_LPSV_QSIS_DELT_250} (d) and Fig. \ref{1_CP_LPSV_HOMO_DELT_250} (d) respectively.

\subsection{One Step Chirping Method : Honeycomb Structure (HC) Case}
\label{QSISMS2_One_Step_Method_HC}

In this case, we have targeted to perform numerical investigations for response of the plasma to the downward chirp in the smaller frequency region and study the formation of the PSV/holes in the respective phase space domains. In one of the previous studies with homogeneous plasmas, Authors \cite{pallavi2016,pallavi2017} reported about the formation and growth of the multiple PSV at different regions of the phase space giving a ``Honeycomb (HC)" like transient structure. In this case, we wanted to understand the fate of such low chirped frequency sweeps formed PSVs in the presence of an background QSIS inhomogeneity. In order to do so, we have excited the plasma with a chirp drive of amplitude $E_{c}^{D}=0.025$ with frequency swept from $\omega_{2}=0.8$ to $\omega_{1}=0.4$ where the sweep rate is $\alpha=-2.0 \times 10^{-3}$ [i.e the chirp coefficent]. Also, the chirp is applied right from the beginning at $t=0~\omega_{pe}^{-1}$ [for homogeneous case] and at $t=120000~\omega_{pe}^{-1}$ [for QSIS inhomogeneous case]. First, we study one case with chirp interval of $\Delta t_{Chirp}=400~\omega_{pe}^{-1}$ with and without ion scale inhomogeneity. In the later section, we increase the chirp interval $\Delta t_{Chirp}$ from 100 to 600 and observe its effect on phase space dynamics.

\begin{figure}
\centerline{\includegraphics[scale=0.50]{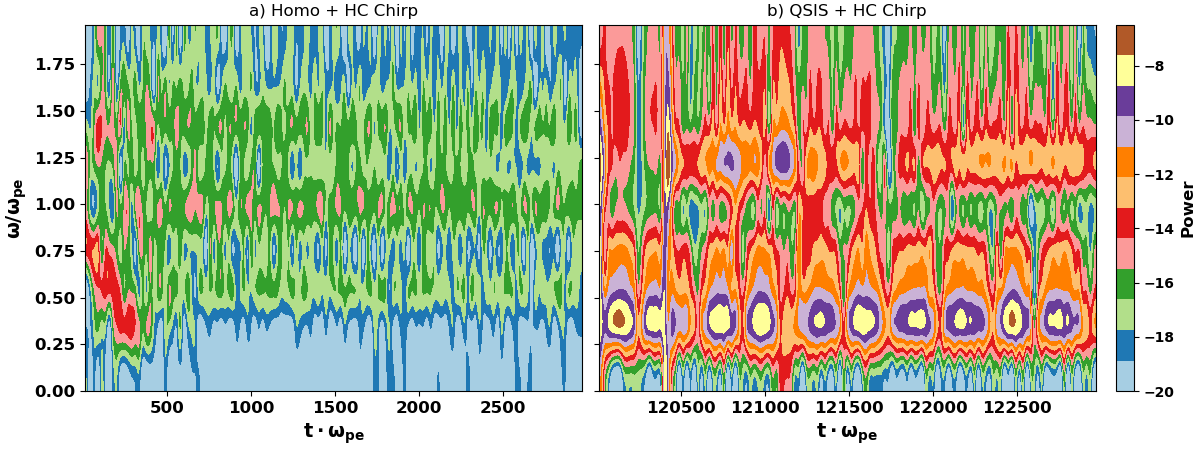}}
\caption{Spectogram plot for (a) Homogeneous + HC Chirp and (b) QSIS inhomogeneity + HC Chirp driven plasma cases with inhomogeneity scale $k_{eq}/k_{min}=2$, chirp interval $\Delta t_{Chirp}=400 ~\omega_{pe}^{-1}$, $E_{c}^{D}=0.025$, chirp drive scale $k_{c}/k_{min}=1$. From (a) and (b), we can observe the generation of various frequency bands which corresponds to different vortex structures in the electron phase space.}
\label{1_SPECTOGRAM_HC}
\end{figure}

Fig. \ref{1_SPECTOGRAM_HC} shows spectogram plot for (a) Homogeneous + HC Chirp and (b) QSIS inhomogeneity + HC Chirp driven plasma cases with inhomogeneity scale $k_{eq}/k_{min}=2$, chirp interval $\Delta t_{Chirp}=400 ~\omega_{pe}^{-1}$, $E_{c}^{D}=0.025$, chirp drive scale $k_{c}/k_{min}=1$. It is quite obvious from the spectogram that Fig.  \ref{1_SPECTOGRAM_HC} (b) is almost analogous to the QSIS inhomogeneity equilibrium construction spectogram upto $t=120000~\omega_{pe}^{-1}$ [See companion paper Part I, Fig. 16 (c)]. It signifies that after the implementation of the HC chirp drive applied between $120000~\omega_{pe}^{-1}<t<120400~\omega_{pe}^{-1}$, no new frequencies are generated in the QSIS inhomogeneity case. Meanwhile for the homogeneous case, we can observe a spread of frequencies in the respective spectogram as shown in Fig. \ref{1_SPECTOGRAM_HC} (a).

\begin{figure}
\centerline{\includegraphics[scale=0.80]{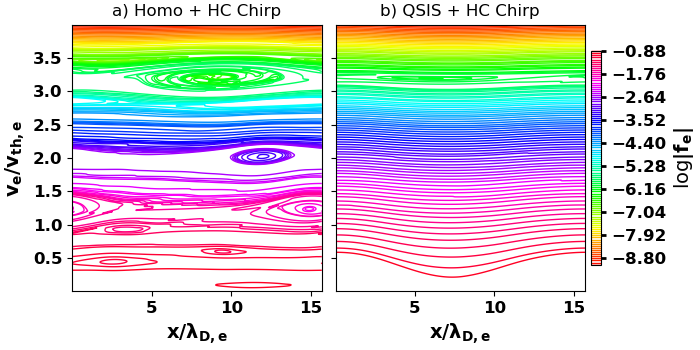}}
\caption{Phase space portrait of late time electron distribution function $f_{e}(x,v)$ for (a) Homogeneous + HC chirp and (b) QSIS inhomogeneity + HC Chirp cases with inhomogeneity scale $k_{eq}/k_{min}=2$, chirp interval $\Delta t_{Chirp}=400 ~\omega_{pe}^{-1}$, $E_{c}^{D}=0.025$, chirp drive scale $k_{c}/k_{min}=1$. Fig. (a) and (b) shows the effect of background QSIS inhomogeneity on the downward HC chirp induced particle trapping.}
\label{1_CP_HC_HOMO+QSIS_DELT_400}
\end{figure}

In Fig. \ref{1_CP_HC_HOMO+QSIS_DELT_400} Phase space portrait of late time electron distribution function $f_{e}(x,v)$ for (a) Homogeneous + HC chirp and (b) QSIS inhomogeneity + HC Chirp cases with inhomogeneity scale $k_{eq}/k_{min}=2$, chirp interval $\Delta t_{Chirp}=400 ~\omega_{pe}^{-1}$, $E_{c}^{D}=0.025$, chirp drive scale $k_{c}/k_{min}=1$. From Fig. \ref{1_CP_HC_HOMO+QSIS_DELT_400} (a), for the homogeneous + HC driven case, we see the formation of multiple PSV corresponding to phase velocities $v_{\phi}^{HC}=3.21,~2.0,~1.25,~1.0,~0.5$. It is evident that the downward chirp HC drive seems to drive the entire sub-harmonic region of the phase space. Closer inspection of phase space reveals that during the driving process, only the large density fluctuations are visible, but at late times smaller PSVs becomes more prominent. The excitations at various phase space velocities gives the distribution a ``Honeycomb (HC)" like appearance. Typically, after the linear drive is switched off, these multi-extrema vortex structures in the sub-harmonic region or HC structures interact continuously, which leads to the merging of the honeycomb vortex [HCV] structures [as expected in the 2D inverse cascading process] resulting into quasi-steady phase space structures in the homogeneous case as shown in Fig. \ref{1_CP_HC_HOMO+QSIS_DELT_400} (a). Surprisingly, with the QSIS inhomogeneous background there is absence of the multi-extrema HCV structures as can be seen in Fig. \ref{1_CP_HC_HOMO+QSIS_DELT_400} (b). The only reason behind the vanishing of the HCV structures is the interaction of these HCV structures with the background QSIS inhomogeneity which either suppresses the formation of these multi-extremas in the phase space or enhances inverse cascading process leading to the asymptotic decrease in the trapped particle fraction to zero simultaneously as the simulation progresses resulting into a phase space with no vortex structures as shown in Fig. \ref{1_CP_HC_HOMO+QSIS_DELT_400} (b).

\subsection{One Step Chirping Method : Response of the system to various Chirp intervals }
\label{QSISMS2_One_Step_Method_Chirp_Intervals}

\begin{figure}
\centerline{\includegraphics[scale=0.63]{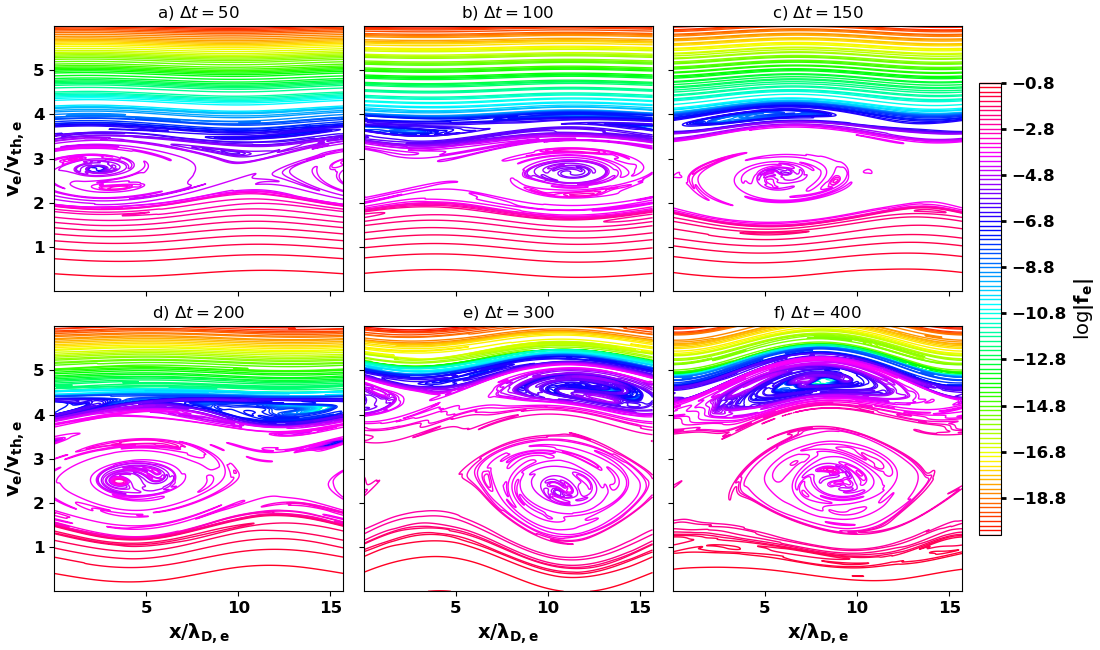}}
\caption{Phase space portrait of electron distribution function $f_{e}(x,v,t=2000~\omega_{pe}^{-1})$ at different chirping intervals i.e (a) $\Delta t_{Chirp}=50~\omega_{pe}^{-1}$, (b) $\Delta t_{Chirp}=100~\omega_{pe}^{-1}$, (c) $\Delta t_{Chirp}=150~\omega_{pe}^{-1}$, (d) $\Delta t_{Chirp}=200~\omega_{pe}^{-1}$, (e) $\Delta t_{Chirp}=300~\omega_{pe}^{-1}$ and (f) $\Delta t_{Chirp}=400~\omega_{pe}^{-1}$ for Homogeneous + LPSV Chirp case with $k_{c}/k_{min}=1$, $E_{c}^{D}=0.025$. Fig. (a) to (f) indicates the increase in the particle trapping leading to enhanced PSVs due to increase in the downward chirping interval $\Delta t_{Chirp}$.}
\label{1_CP_LPSV_HOMO_VARIOUS_DELT}
\end{figure}

\begin{figure}
\centerline{\includegraphics[scale=0.63]{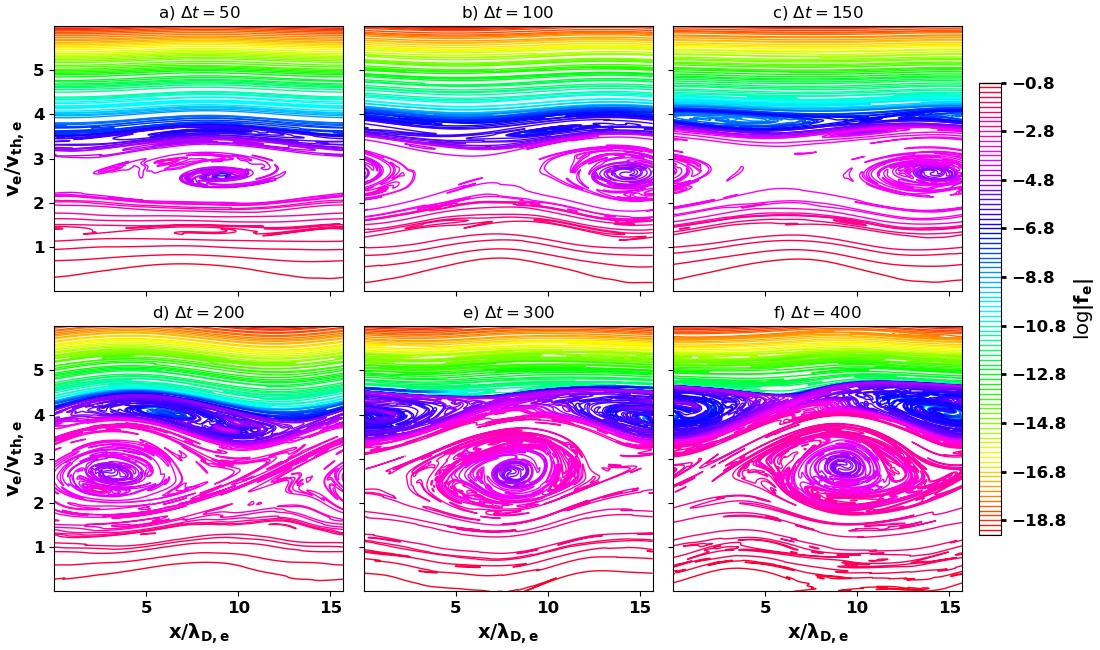}}
\caption{Phase space portrait of electron distribution function $f_{e}(x,v,t=120000~\omega_{pe}^{-1})$ at different chirping intervals i.e (a) $\Delta t_{Chirp}=50~\omega_{pe}^{-1}$, (b) $\Delta t_{Chirp}=100~\omega_{pe}^{-1}$, (c) $\Delta t_{Chirp}=150~\omega_{pe}^{-1}$, (d) $\Delta t_{Chirp}=200~\omega_{pe}^{-1}$, (e) $\Delta t_{Chirp}=300~\omega_{pe}^{-1}$ and (f) $\Delta t_{Chirp}=400~\omega_{pe}^{-1}$ for QSIS Inhomogeneity + LPSV Chirp case with inhomogeneity scale $k_{eq}/k_{min}=2$, $E_{c}^{D}=0.025$, chirp drive scale $k_{c}/k_{min}=1$. Fig. (a) to (f) indicates the increase in the particle trapping leading to enhanced PSVs due to increase in the downward chirping interval $\Delta t_{Chirp}$. Additionaly, it also illustrates the supression of PSVs in the presence of background QSIS inhomogeneity.}
\label{1_CP_LPSV_QSIS_VARIOUS_DELT}
\end{figure}

As mentioned earlier, in this section, we will present the response of the chirp driven plasma systems to increase in the LPSV/HC chirp intervals with background QSIS inhomogeneity. Individual electron phase space evolution is also compared to their homogeneous counterparts. Fig. \ref{1_CP_LPSV_HOMO_VARIOUS_DELT} and Fig. \ref{1_CP_LPSV_QSIS_VARIOUS_DELT} demonstrates phase space portrait of electron distribution functions  i.e $f_{e}(x,v,t=2000~\omega_{pe}^{-1} - 122000~\omega_{pe}^{-1})$ at different chirping intervals i.e (a) $\Delta t_{Chirp}=50~\omega_{pe}^{-1}$, (b) $\Delta t_{Chirp}=100~\omega_{pe}^{-1}$, (c) $\Delta t_{Chirp}=150~\omega_{pe}^{-1}$, (d) $\Delta t_{Chirp}=200~\omega_{pe}^{-1}$, (e) $\Delta t_{Chirp}=300~\omega_{pe}^{-1}$ and (f) $\Delta t_{Chirp}=400~\omega_{pe}^{-1}$ for Homogeneous + LPSV Chirp and QSIS inhomogeneity + LPSV cases with $k_{c}/k_{min}=1$, $E_{c}^{D}=0.025$. It is evident from Fig. \ref{1_CP_LPSV_HOMO_VARIOUS_DELT} [(a) to (f)], that with the increase in the chirp interval $\Delta t_{Chirp}$ from 50 to 400, the size of the LPSV structures formed in the electron phase space increases due to increase in the trapped particle fractions. Moreover, at higher amplitude of chirp intervals i.e $\Delta t_{Chirp}=300$ or $400~\omega_{pe}^{-1}$ more particles are trapped in the LPSV structure corresponding to the high phase velocitiy which is in contrast to the low amplitude chirp intervals. Similar to the homogeneous case, with the increase in the chirp interval $\Delta t_{Chirp}$ from 50 to 400, we have seen the increase in the size of the LPSV structures in the electron phase space. However, on the contrary if one compares Fig. \ref{1_CP_LPSV_HOMO_VARIOUS_DELT} [(a) to (f)] and Fig. \ref{1_CP_LPSV_QSIS_VARIOUS_DELT} [(a) to (f)], it is obvious that there is some degree of supression in the particle trapping phenomenon in the QSIS inhomogeneity case than homogeneous case and because of which the size of the LPSV structures of QSIS case are visually smaller compared to the LPSV structures in the homogeneous case.

\begin{figure}
\centerline{\includegraphics[scale=0.045]{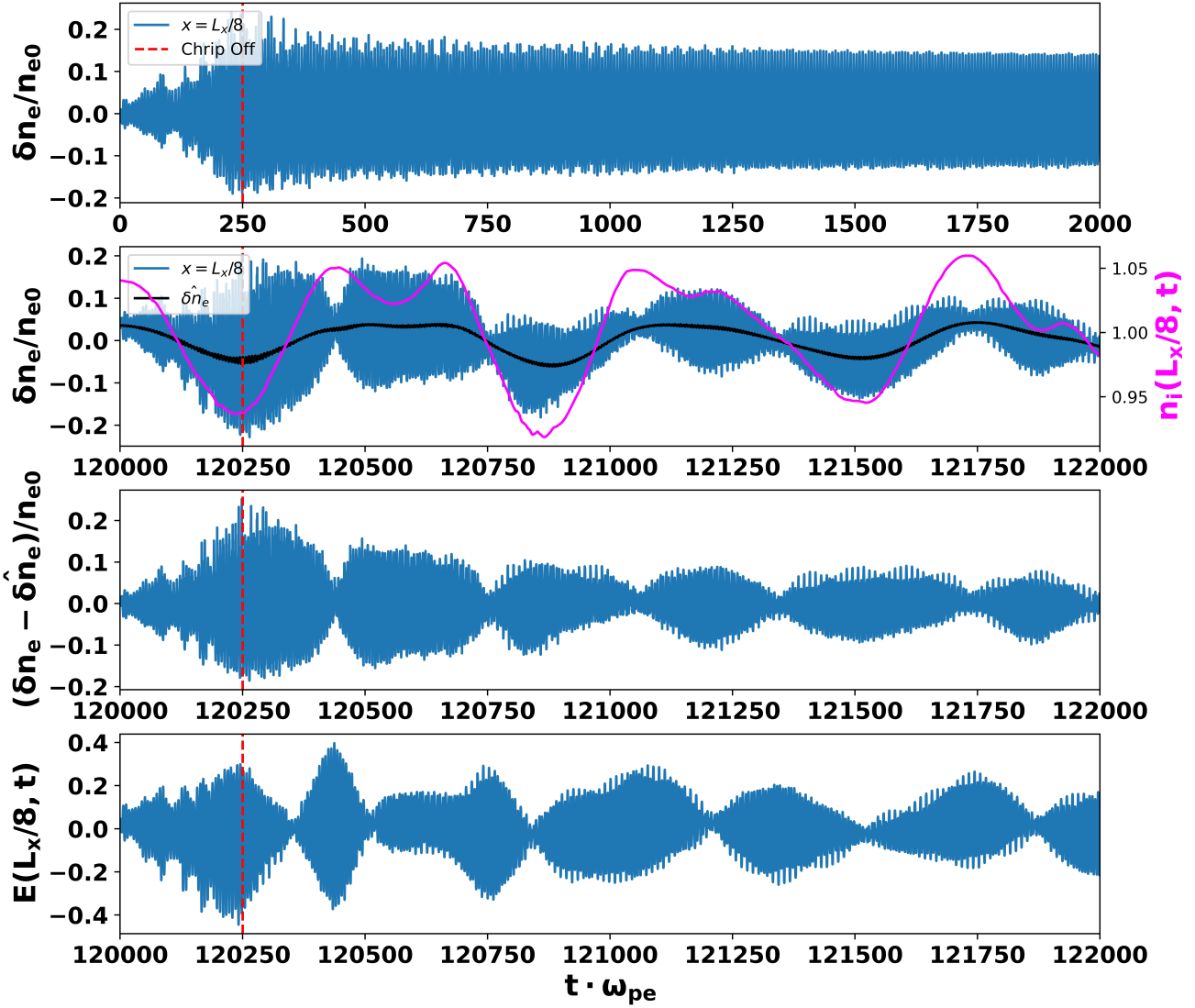}}
\caption{Temporal evolution of (a) homogeneous case excess density fraction (EDF) of electrons i.e $\delta n_{e}/n_{e0}$ at $x=L_{max}/8$, (b) electron EDF at $x=L_{max}/8$, ion density $n_{i}(x=L_{max}/8,t)$ and moving average variation $\delta \hat{n_{e}}$, (c) difference of electron EDF and moving average values i.e $(\delta n_{e} - \delta \hat{n_{e}})/n_{e0}$ at $x=L_{max}/8$, (d) electric field variation at $x=L_{max}/8$ for both homogeneous and QSIS inhomogeneity + LPSV Chirp driven plasma cases with $k_{eq}/k_{min}=2$, $E_{c}^{D}=0.025$, $k_{c}/k_{min}=1$ and chirp interval $\Delta t_{Chirp}=250 ~\omega_{pe}^{-1}$ respectively. In the legend, the dotted line denote the switch off time of LPSV chirp drive as denoted by $\psi_{1}$ in Table \ref{TABLE_2}. }
\label{1_EDF_LPSV}
\end{figure}

In order to get some qualitative estimate to support our argument about PSV structure sizes and trapping and untrapping dynamics, we have used the similar approach and diagnostics as discussed in the Sec. \ref{QSISMS2_Two_Step_Method} for SEAW perturbation case. In Table \ref{TABLE_2}, we have tabulated the electron excess density fraction [EDF : $\delta n_{e}/n_{e0}$, defined in the companion paper Part I] for the homogeneous case alongwith the difference of electron EDF and moving average $\delta \hat{n_{e}}$ value i.e $(\delta n_{e}-\delta \hat{n_{e}})/n_{e0}$ for the QSIS inhomogeneity case at two temporal locations denoted by $T_{1}$ and $T_{2}$ respectively, where, $T_{1}$ is the end of the LPSV downward chirped frequency drive [$t=250~ or ~120250~\omega_{pe}^{-1}$] and $T_{2}$ is the end of the simulation time [$t=2000~ or ~122000~\omega_{pe}^{-1}$] for respective cases. Alongwith it just to set the reference for an individual LPSV chirp case, Fig. \ref{1_EDF_LPSV} demonstrates temporal variation of (a) homogeneous case excess density fraction (EDF) of electrons i.e $\delta n_{e}/n_{e0}$ at $x=L_{max}/8$, (b) electron EDF at $x=L_{max}/8$, ion density $n_{i}(x=L_{max}/8,t)$ and moving average variation $\delta \hat{n_{e}}$, (c) difference of electron EDF and moving average values i.e $(\delta n_{e} - \delta \hat{n_{e}})/n_{e0}$ at $x=L_{max}/8$, (d) electric field variation at $x=L_{max}/8$ for both homogeneous and QSIS inhomogeneity + LPSV Chirp driven plasma cases with $k_{eq}/k_{min}=2$, $E_{c}^{D}=0.025$, $k_{c}/k_{min}=1$ and chirp interval $\Delta t_{Chirp}=250 ~\omega_{pe}^{-1}$ respectively. The dotted line denote the LPSV chirp drive switch off time as also denoted by $T_{1}$ in Table \ref{TABLE_2}.

\begin{table}
\caption []{Electron excess density fraction [EDF : $(\delta n/n_{0})$] for homogeneous and $(\delta n_{e}-\delta \hat{n_{e}})/n_{e0}$ for the QSIS inhomogeneity case at $x=L_{max}/8$ estimated at various temporal locations denoted by $\psi_{1}$ and $\psi_{2}$ respectively (as shown in Fig. \ref{1_EDF_LPSV}), with $k_{eq}/k_{min}=2$, $k_{c}/k_{min}=1$, $E_{c}^{D}=0.025$ and chirp interval $\Delta t_{Chirp}=250 ~\omega_{pe}^{-1}$.}   
\centering                         
\begin{tabular}{c c c c c c c c c}           
\hline\hline                        
Case & Position &  50  &  100  & 150 & 200 & 250 & 300 & 400   \\ [1.0ex]    
\hline          
Homogeneous & $T_{1}$  & 0.058 & 0.056 & 0.074 & 0.162 & 0.242 & 0.380 & 0.22 \\          
              & $T_{2}$  & 0.024 & 0.025 & 0.041 & 0.061 & 0.140 & 0.132 & 0.134 \\          
QSIS Inhomogeneity & $T_{1}$ & 0.049 & 0.049 & 0.078 & 0.178 & 0.244 & 0.30 & 0.191 \\          
              & $T_{2}$ & 0.020 & 0.0169 & 0.032 & 0.062 & 0.032 & 0.092 & 0.09 \\ [1ex]
\hline\hline                               
\end{tabular}
\label{TABLE_2}
\end{table}

In Fig. \ref{1_EDF_LPSV} (a) and from Table \ref{TABLE_2}, we can observe for the homogeneous case with chirp interval $\Delta t_{Chirp}=250 ~\omega_{pe}^{-1}$, that during the LPSV chirp drive, the electron EDF increases from zero to 0.242 value rapidly in the interval of $0~\omega_{pe}^{-1} <t < 250~\omega_{pe}^{-1}$ infering particle trapping phenomenon. Afterwards, when the drive is switched off i.e $t> 250~\omega_{pe}^{-1}$, the electron EDF relaxes to a finite oscillatory non-zero value of 0.140 i.e EDF$|_{T_{2}}=$0.140 at $T_{2}=2000~\omega_{pe}^{-1}$. Also, from Table \ref{TABLE_2}, we can refer the EDF amplitude variation throughout $T_{1}$ and $T_{2}$ temporal locations respectively, suggesting, that for $\Delta t_{Chirp}=250~\omega_{pe}^{-1}$ LPSV homogeneous case, the particle trapping was at its peak just after the LPSV chirp drive was switched off at $t=250~\omega_{pe}^{-1}$. In the same way, for the QSIS inhomogeneity case, between $120000~\omega_{pe}^{-1} <t < 120250~\omega_{pe}^{-1}$, analogous particle trapping trend can be seen from Fig. \ref{1_EDF_LPSV} (c) i.e increase in the $(\delta n_{e}-\delta \hat{n_{e}})/n_{e0}$ value during the applied LPSV chirp drive. On the contrary to the homogeneous case, after the LPSV drive is switched off,  $(\delta n_{e}-\delta \hat{n_{e}})/n_{e0}$ varies as collection of small pockets illustrating simultaneous particle trapping/untrapping dynamics in the electron phase space. At $T_{2}$ temporal location, we have observed greater relaxation in the QSIS EDF estimates when compared to the homogeneous case i.e EDF$|_{T_{2}}^{Homo}=0.140$ and EDF$|_{T_{2}}^{QSIS}=0.032$ supporting our argument of suppression of PSV formation in the electron phase space due to enhanced late time particle untrapping in the presence of background QSIS inhomogeneity. In addition, from Fig. \ref{1_EDF_LPSV} (d), one can see the temporal electric field variation around $x=L_{max}/8$ for the QSIS inhomogeneity case showing the rapid increase in the field amplitude during downward LPSV chirp drive $120000~\omega_{pe}^{-1} <t < 120250~\omega_{pe}^{-1}$ as well as oscillatory pockets of amplitude variation after the drive is switched off i.e $t>120250~\omega_{pe}^{-1}$. Late time finite non-zero electric field amplitude at $T_{2}$ confirms that the obtained solution is a BGK mode.

\begin{figure}
\centerline{\includegraphics[scale=0.27]{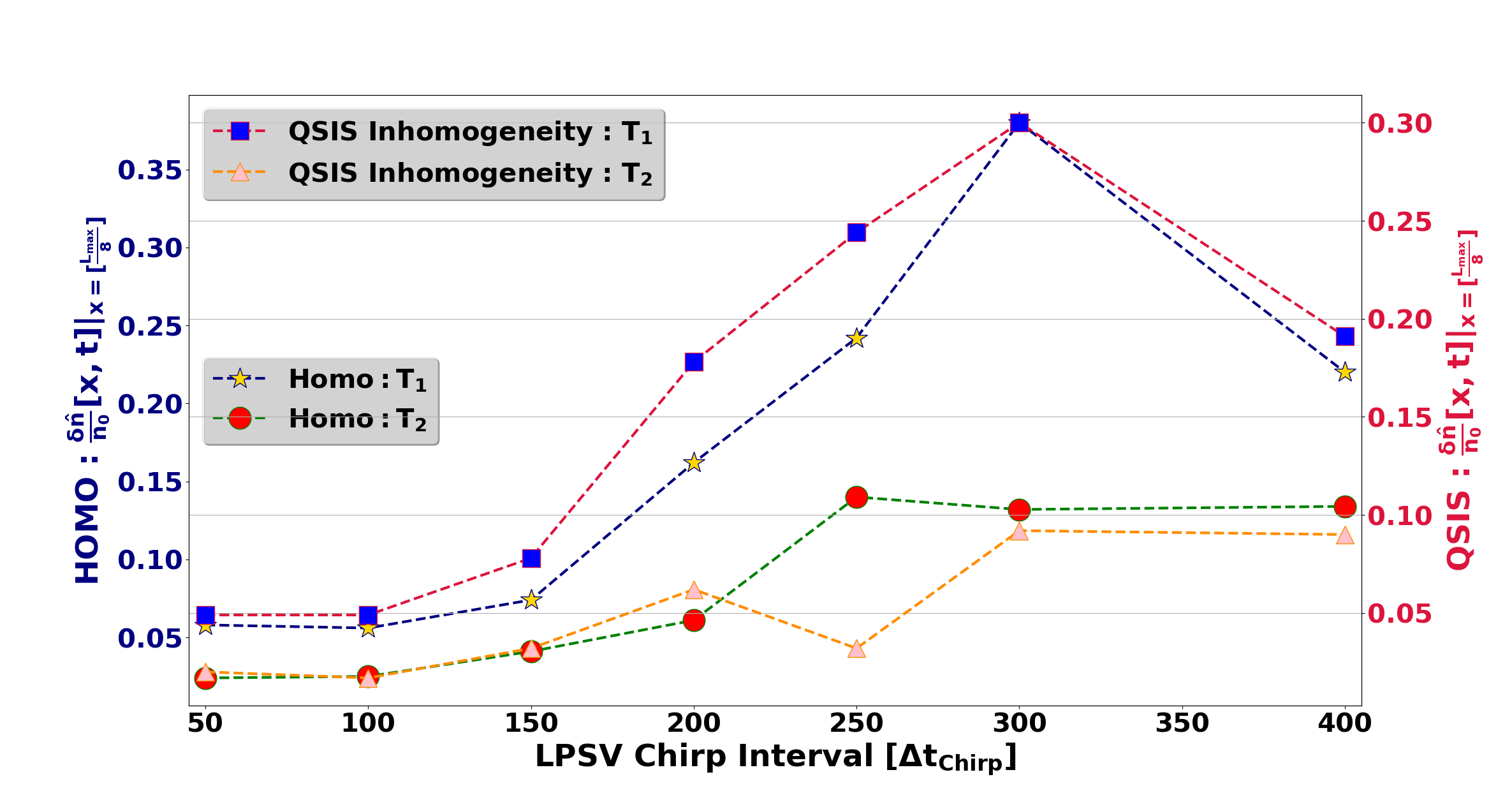}}
\caption{Variation of the difference of electron EDF and moving average values i.e EDF : $(\delta n_{e} - \delta \hat{n_{e}})/n_{e0}$ at $x=L_{max}/8$ for LPSV + QSIS inhomogeneous case and EDF : $\delta n_{e}/n_{e0}$ at $x=L_{max}/8$ for LPSV + homogeneous case with respect to LPSV chirp intervals $\Delta t_{Chirp}$ with $k_{eq}/k_{min}=2$, $k_{c}/k_{min}=1$, and $E_{c}^{D}=0.025$. Here, temporal locations $T_{1}$ and $T_{2}$ denotes the LPSV chirp switch off time and end of the simulation time respectively. The data plotted here is tabulated in Table \ref{TABLE_2}.}
\label{1_EDF_VS_CHIRP_LPSV}
\end{figure}

Fig. \ref{1_EDF_VS_CHIRP_LPSV} illustrates the variation of EDF : $(\delta n_{e} - \delta \hat{n_{e}})/n_{e0}$ values at $x=L_{max}/8$ for LPSV + QSIS inhomogeneous case and EDF : $\delta n_{e}/n_{e0}$ at $x=L_{max}/8$ for LPSV + homogeneous case with respect to chirp intervals $\Delta t_{Chirp}$ ranging from 50 to 400 with $k_{eq}/k_{min}=2$, $k_{c}/k_{min}=1$, and $E_{c}^{D}=0.025$ as tabulated in Table \ref{TABLE_2}. Here, temporal locations $T_{1}$ and $T_{2}$ denotes the LPSV chirp switch off time and end of the simulation time respectively. It is obvious from Fig. \ref{1_EDF_VS_CHIRP_LPSV} that the EDF estimates at the temporal location $T_{1}$ is greater than those at location $T_{2}$ for both the homogeneous and inhomogeneous cases. Evidently, at both the $T_{1}$ and $T_{2}$ locations the increase in the EDF estimates for both cases are not monotonic i.e  one can not always expect that increase in the chirp interval $\Delta t_{Chirp}$ will always result into increased EDF or more particle trapping phenomenon. For example, Fig. \ref{1_EDF_VS_CHIRP_LPSV} reflects that at location $T_{1}$, for both homogeneous and QSIS inhomogeneous cases, there is continuous increase in the EDF value with the increase in LPSV $\Delta t_{Chirp}$ till $\Delta t_{Chirp}=300~\omega_{pe}^{-1}$, at the later value EDF estimate falls down despite greater chirp interval. Similarly, at location $T_{2}$, the increase in continuous till $\Delta t_{Chirp}=200~\omega_{pe}^{-1}$, later on, the EDF values are non-monotonic in nature. One more interesting observation is that in the presence of QSIS inhomogeneous background EDF values either reduces or becomes equal to their homogeneous counterparts for all the chirp intervals ranging from 50 to 400 leading to reduced PSV structure sizes as seen in Figs, \ref{1_CP_LPSV_HOMO_VARIOUS_DELT} and \ref{1_CP_LPSV_QSIS_VARIOUS_DELT} respectively. Not even in one single case studied in the present work this QSIS inhomogeneity enhances the EDF estimates.

\begin{figure}
\centerline{\includegraphics[scale=0.63]{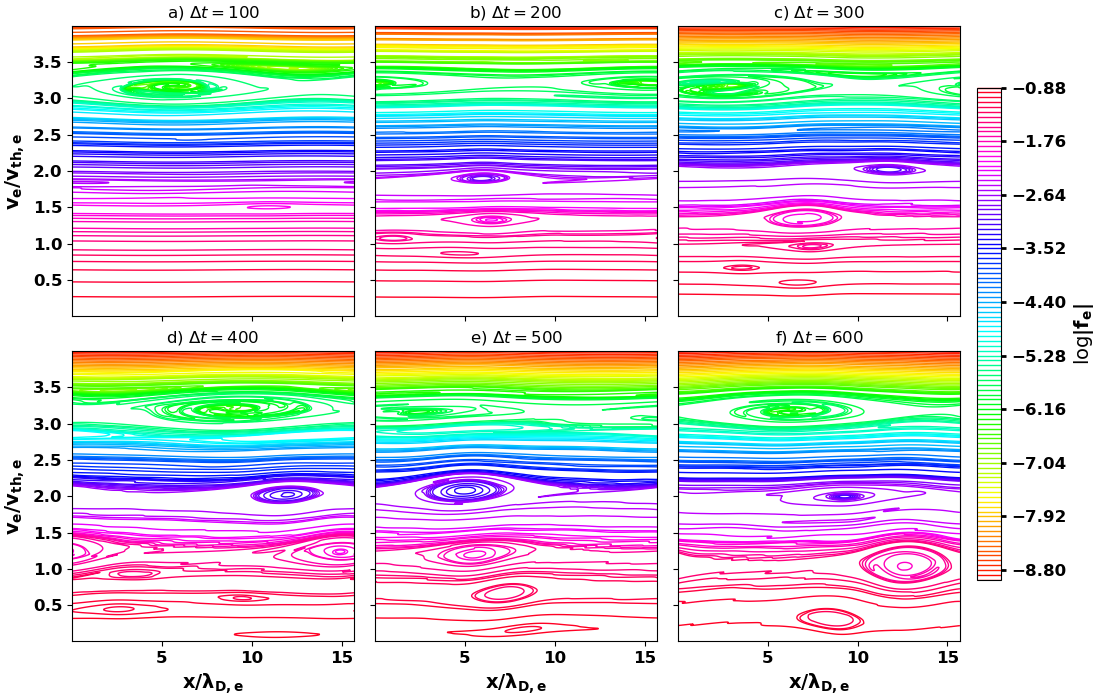}}
\caption{Phase space portrait of electron distribution function $f_{e}(x,v,t=3000~\omega_{pe}^{-1})$ at different chirping intervals i.e (a) $\Delta t_{Chirp}=100~\omega_{pe}^{-1}$, (b) $\Delta t_{Chirp}=200~\omega_{pe}^{-1}$, (c) $\Delta t_{Chirp}=300~\omega_{pe}^{-1}$, (d) $\Delta t_{Chirp}=400~\omega_{pe}^{-1}$, (e) $\Delta t_{Chirp}=500~\omega_{pe}^{-1}$ and (f) $\Delta t_{Chirp}=600~\omega_{pe}^{-1}$ for Homogeneous + HC Chirp case with $E_{c}^{D}=0.025$, chirp drive scale $k_{c}/k_{min}=1$. Fig. (a) to (f) illustrates the formation of various vortex structures alongwith enhanced particle trapping with the increase in the chirping interval $\Delta t_{Chirp}$ from $100$ to $600$. }
\label{1_CP_HC_HOMO_VARIOUS_DELT}
\end{figure}

\begin{figure}
\centerline{\includegraphics[scale=0.63]{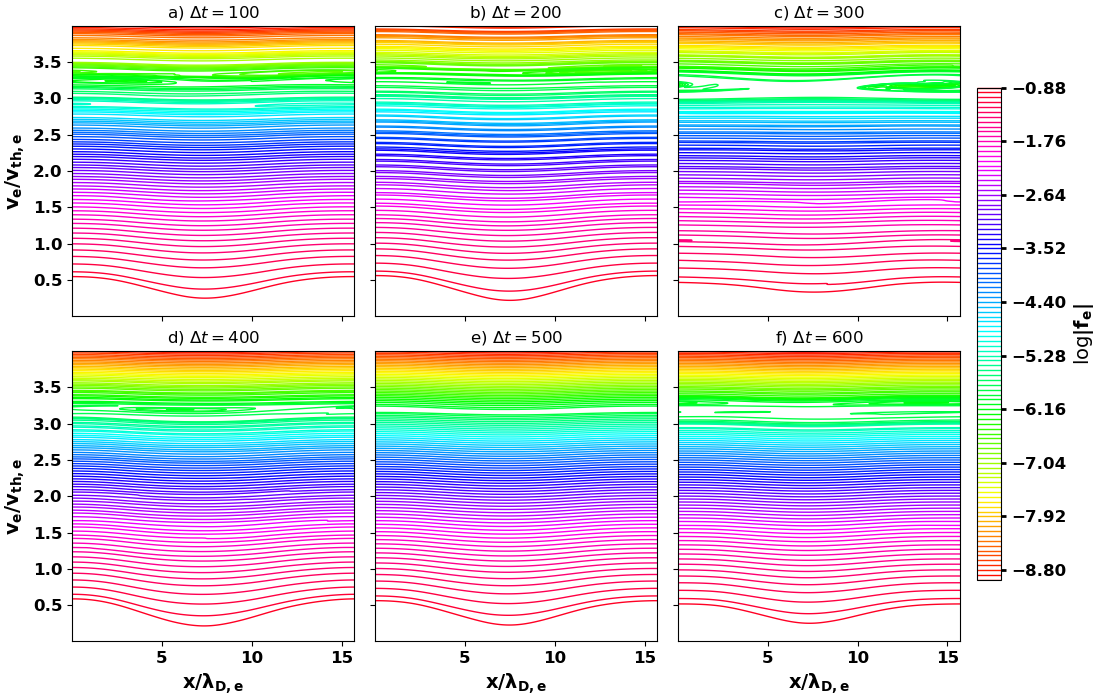}}
\caption{Phase space portrait of electron distribution function $f_{e}(x,v,t=123000~\omega_{pe}^{-1})$ at different chirping intervals i.e (a) $\Delta t_{Chirp}=100~\omega_{pe}^{-1}$, (b) $\Delta t_{Chirp}=200~\omega_{pe}^{-1}$, (c) $\Delta t_{Chirp}=300~\omega_{pe}^{-1}$, (d) $\Delta t_{Chirp}=400~\omega_{pe}^{-1}$, (e) $\Delta t_{Chirp}=500~\omega_{pe}^{-1}$ and (f) $\Delta t_{Chirp}=600~\omega_{pe}^{-1}$ for QSIS Inhomogeneity + HC Chirp case with inhomogeneity scale $k_{eq}/k_{min}=2$, $E_{c}^{D}=0.025$, chirp drive scale $k_{c}/k_{min}=1$. Fig. (a) to (f) indicates that there is no major effect of chirp driven particle trapping due to the presence of background QSIS inhomogeneity. }
\label{1_CP_HC_QSIS_VARIOUS_DELT}
\end{figure}

Fig. \ref{1_CP_HC_HOMO_VARIOUS_DELT} and Fig. \ref{1_CP_HC_QSIS_VARIOUS_DELT} illustrates Phase space portrait of electron distribution functions i.e $f_{e}(x,v,t=3000~\omega_{pe}^{-1} - 123000~\omega_{pe}^{-1})$ for homogeneous + HC chirp and QSIS inhomogeneous + HC chirp cases at different chirping intervals i.e (a) $\Delta t_{Chirp}=100~\omega_{pe}^{-1}$, (b) $\Delta t_{Chirp}=200~\omega_{pe}^{-1}$, (c) $\Delta t_{Chirp}=300~\omega_{pe}^{-1}$, (d) $\Delta t_{Chirp}=400~\omega_{pe}^{-1}$, (e) $\Delta t_{Chirp}=500~\omega_{pe}^{-1}$ and (f) $\Delta t_{Chirp}=600~\omega_{pe}^{-1}$ with inhomogeneity scale $k_{eq}/k_{min}=2$, $E_{c}^{D}=0.025$, HC chirp drive scale $k_{c}/k_{min}=1$. From Fig. \ref{1_CP_HC_HOMO_VARIOUS_DELT} [(a) to (f)], we can clearly observe that with the increase in the HC drive chirp interval $\Delta t_{Chirp}$ from 100 to 600, the size of the multi-extrema HCV structures become more prominent with large trapped particles in the homogeneous background case. Also, large chirping interval excites more number of HCV structures in the sub-harmonic region as compared to the low values of the chirping interval [See. Fig. \ref{1_CP_HC_HOMO_VARIOUS_DELT} (a) to (f)]. Meanwhile, in contrast to the homogeneous case, there ia an absence of any HCV structures in the phase space of QSIS inhomogeneity case even though we have increased the $\Delta t_{Chirp}$ from 100 to 600, indicating that the low frequency sweeps are inefficient to form any kind of particle trapping structures in phase space as shown in Fig. \ref{1_CP_HC_QSIS_VARIOUS_DELT} [(a) to (f)] respectively.

\begin{figure}
\centerline{\includegraphics[scale=0.045]{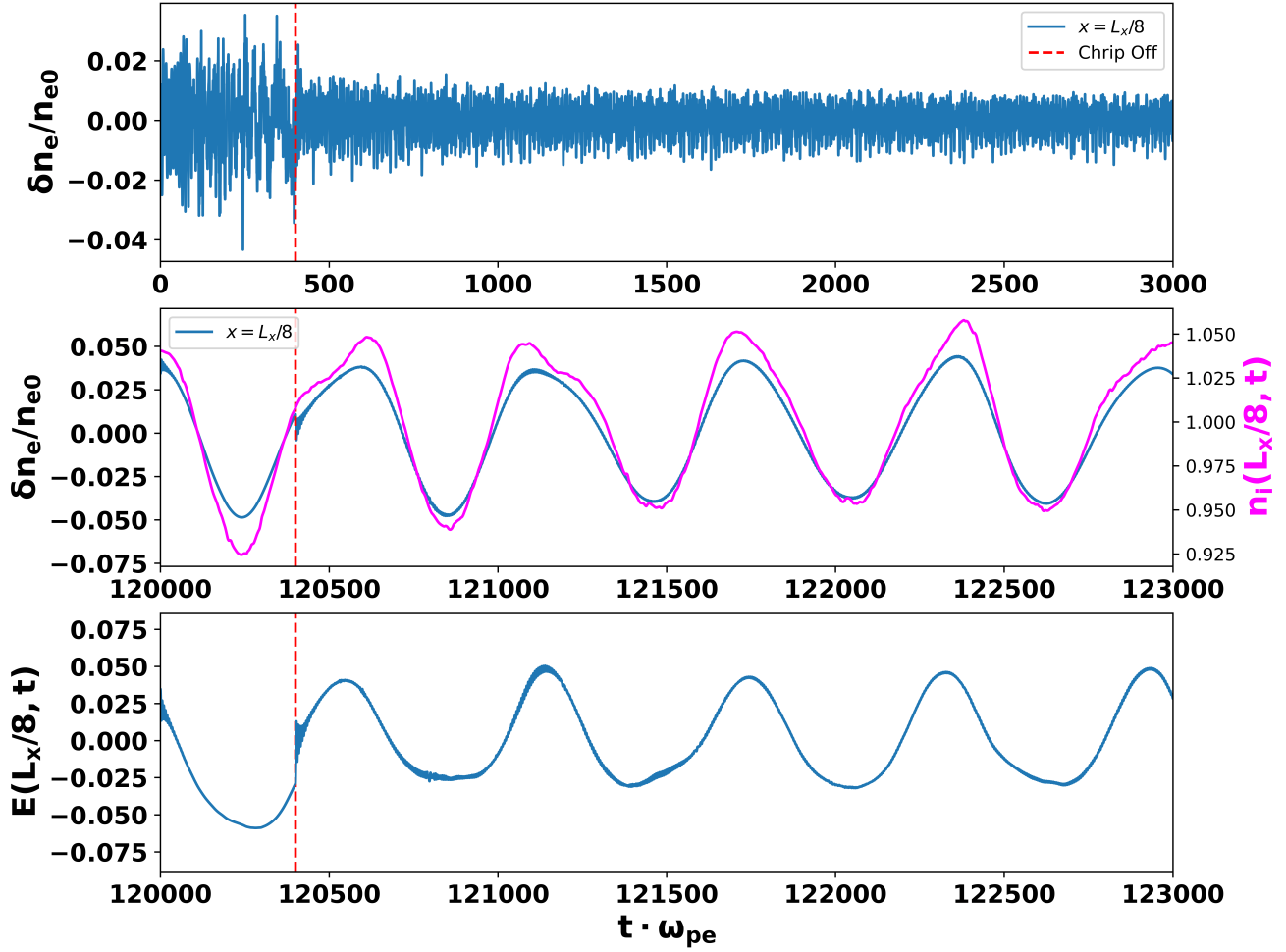}}
\caption{Temporal evolution of (a) homogeneous case excess density fraction (EDF) of electrons i.e $\delta n_{e}/n_{e0}$ at $x=L_{max}/8$, (b) electron EDF at $x=L_{max}/8$, ion density $n_{i}(x=L_{max}/8,t)$, (c) electric field variation at $x=L_{max}/8$ for both homogeneous and QSIS inhomogeneity + HC Chirp driven plasma cases with $k_{eq}/k_{min}=2$, $E_{c}^{D}=0.025$, $k_{c}/k_{min}=1$ and chirp interval $\Delta t_{Chirp}=400 ~\omega_{pe}^{-1}$ respectively. In the legend, the dotted line denote the switch off time of HC chirp drive.}
\label{1_EDF_HC}
\end{figure}

In Fig. \ref{1_EDF_HC}, we demonstrate temporal evolution of (a) homogeneous case excess density fraction (EDF) of electrons i.e $\delta n_{e}/n_{e0}$ at $x=L_{max}/8$, (b) electron EDF at $x=L_{max}/8$, ion density $n_{i}(x=L_{max}/8,t)$, (c) electric field variation at $x=L_{max}/8$ for both homogeneous and QSIS inhomogeneity + HC Chirp driven plasma cases with $k_{eq}/k_{min}=2$, $E_{c}^{D}=0.025$, $k_{c}/k_{min}=1$ and chirp interval $\Delta t_{Chirp}=400 ~\omega_{pe}^{-1}$ respectively. It reflects that for homogeneous case we can observe some finite non-zero EDF value which suggest the existence of PSV structures in the electron phase space [See Fig. \ref{1_CP_HC_HOMO+QSIS_DELT_400}]. Meanwhile, for the QSIS inhomogeneous case the electron EDF is heavily influenced by the presence of a strong ion background leading to the absence of PSV structures. As Fig. \ref{1_EDF_HC} (b) indicates, we were also unable to find the moving average $\hat{\delta n_{e}}$ of the electron EDF contrary to the previous SEAW and LPSV chirp driven cases. Also, at $x=L_{max}/8$ we observe a oscillatory finite non-zero temporal variation of the electric field as shown in Fig. \ref{1_EDF_HC} (c). In all of the cases present in this work, energy conservation and corresponding electron and ion numerical entropy conservation [defined in the companion paper Part I] holds good for the given grid resolution indicating absence of any inconsistency in the simulations due to numerical artifacts.

\section{Discussion and Conclusion}
\label{QSISMS2_Discussion_conclusion}

In this work i.e Part II, using high resolution long time Vlasov-Poisson simulations with kinetic ions and kinetic electrons, we have investigated the response of various PSVs generated using two step [SEAW case] or one step [LPSV or HC case] downward chirp frequency drive in the presence of a QSIS inhomogeneity obtained in the companion paper i.e Part I. Also, for the exact simulation parameter sets, individual chirp inhomogeneous cases were compared to their homogeneous counterparts. In the following, we have summarized the important results as follows $\longrightarrow$
\begin{itemize}
	\item In the two step driving method i.e constant frequency EAW drive + downward chirp frequency drive, we have observed the following :
	\begin{itemize}
		\item In the 2D $(\omega,k)$ power spectrum plot [See Fig. \ref{1_2DPS_SEAW}], for the QSIS inhomogeneity case, we have noticed the suppression of frequencies in the range of $\omega/\omega_{pe}=1.0$ to $\omega/\omega_{pe}=2.0$ when compared to the homogeneous case. Also, wave-wave mode coupling phenomenon was more prominent in the QSIS inhomogeneity case compared to the homogeneous case. 
		
		\item The nature of the spectogram generated in the QSIS case is more descrete compared to the homogeneous case [See Fig \ref{1_SPECTOGRAM_SEAW}].
		
		\item We have observed the early onset of Langmuir (LAN) mode alongwith more particle trapping in the separatrix region of the PSV, in the QSIS inhomogeneous case when compared to the homogeneous case. 
		phenomenon
		\item In this case, the phase space structure in the ion phase space i.e QSIS inhomogeneity sustains throughout the simulation without getting affected by any constant frequency perturbation or chirp drives.
		   
	 	\item In the EAW + QSIS inhomogeneous case, we have to find the moving average of the electron excess density fraction $\hat{\delta n_{e}}$ and had to calculate the quantity $(\delta n_{e}-\delta \hat{n_{e}})/n_{e0}$ to get the trapping/untrapping fraction estimates. From Table \ref{TABLE_1}, we can conclude that the particle trapping was at its maximum just after the chirp drive was switched off for both the homogeneous and QSIS inhomogeneous cases.
	 	
	 	\item We have also observed relaxation in the EDF estimates for the QSIS inhomogeneous case when compared to the homogeneous counterpart indicating the supression of PSV formation in the electron phase space [See Figs. \ref{1_CP_SEAW_HOMO} and \ref{1_CP_SEAW_QSIS}].
	\end{itemize}
	
	\item In the one step downward chirped frequency driving method [LPSV or HC case], we have observed the following :
	\begin{itemize}
		\item Apart from presence or absence of wave-wave mode coupling interactions, we have noticed a discrete distribution of power spectrum in 2D $(\omega,k)$ in the range of $\omega/\omega_{pe}=0.75$ to $\omega/\omega_{pe}=2.0$ values for QSIS inhomogeneity + LPSV chirp drive case when compared to their homogeneous counterpart.
		
		\item In the spectogram signature, we have noticed the discontinuity of frequency band around $\omega_{c}=1.0$ in the QSIS inhomogeneous case, whereas, it is continuous in the homogeneous counterpart. Additionally, generation of high amplitude temporal frequency regions can be seen in the QSIS case compared to the homogeneous case [See Fig. \ref{1_SPECTOGRAM_LPSV}].
		
		\item In the LPSV chirp driven case, analogous to the previous case, we have observed more particle trapping in the separatrix region in the QSIS inhomogeneous case compared to the homogeneous case. Also, it is evident that after the relaxation, particle untrapping in the homogeneous case is higher from its counterpart i.e QSIS inhomogeneous case.
		
		\item In the HC chirp drive case, no new frequencies are generated in the spectogram, meanwhile , if we look its counterpart we can observe the spread of frequencies in the spectogram. 
		
		\item Also, for the HC chirp driven case with $\Delta T_{Chirp}=400$, one striking difference exists i.e, we can see the formation of multiple PSV [HCV Structures] corresponding to the phase velocities $v_{\phi}^{HC}=$3.21,2.0,1.25,1.0,0.5 in homogeneous case. But, there is absence of these multi-extrema HCV structures in the QSIS inhomogeneous case [See Fig. \ref{1_CP_HC_HOMO+QSIS_DELT_400}].   
		
	\end{itemize}
	
	\item In the LPSV or HC chirp driven cases, we can conclude that the presence of background QSIS inhomogeneity supresses the formation of corresponding PSV structures in the electron phase space [See Figs. \ref{1_CP_LPSV_HOMO_VARIOUS_DELT}, \ref{1_CP_LPSV_QSIS_VARIOUS_DELT}, \ref{1_CP_HC_HOMO_VARIOUS_DELT}, \ref{1_CP_HC_QSIS_VARIOUS_DELT} and Table \ref{TABLE_2}]. 
	
	\item From Fig. \ref{1_EDF_VS_CHIRP_LPSV}, we can conclude that EDF estimates just after the chirp drive is switched off is greater than those at the end of the simulation for both the cases.
	
	\item With the increase in the LPSV chirp interval, the increase in the EDF estimates are not monotonic in nature for both the QSIS and homogeneous cases respectively [Fig. \ref{1_EDF_VS_CHIRP_LPSV}].
	
	\item Not even in a single case among the cases investigated in this work, we found that QSIS inhomogeneity enhances the EDF estimates. 
	
\end{itemize}

In all of the above simulations runs the energy and entropy conservation holds good for the choosen grid resolution indicating high quality stable solution without any numerical artifact. Our study brings out several key aspects of evolutionary dynamics of chirp driven PSVs in the background of QSIS inhomogeneity which may be crucial in understanding the fundamental phenomenon such as wave-wave mode coupling interaction, wave-particle interactions, collisionless turbulence or multi-mode plasma dynamics relevant to laboratory as well as astrophysical plasmas.


\section*{Acknowledgments}
All the computational results of this paper were obtained using the ANTYA HPC Linux cluster at Institute for Plasma Research (IPR) Gandhinagar, India. The authors would like to thank the Data Center staff at IPR.


\section*{Data availability statement}
The data that support the findings of this study are available upon reasonable request from the authors.

\section*{References}
\bibliography{iopart-num}

\end{document}